\documentclass[a4,10pt,usenatbib,twocolumn]{mn2e}
\usepackage[dvipdf]{graphicx, psfrag}
\usepackage{times}
\usepackage{amsmath}
\usepackage{natbib}
\usepackage{aas_macros}

 \title[Axisymmetric and stationary magnetized barotropic stars]
 {Axisymmetric and stationary structures of magnetized barotropic 
   stars
 with extremely strong magnetic fields deep inside}
\author[K. Fujisawa, S'i. Yoshida and Y. Eriguchi]{Kotaro
Fujisawa
\thanks{E-mail: fujisawa@ea.c.u-tokyo.ac.jp},
Shin'ichirou Yoshida
and Yoshiharu Eriguchi \\
Department of Earth Science and Astronomy,
Graduate School of Arts and Sciences, University of Tokyo,\\
Komaba, Meguro-ku, Tokyo 153-8902, Japan}

\date{Accepted 2012 January 23. Received 2012 January 23; in original form 2010 October 30}

\def\Vec#1{\mbox{\boldmath $#1$}}
\def\D#1#2{\dfrac{d #1}{d #2}}

\def\P#1#2{\dfrac{\partial #1}{\partial #2}}

\begin{document}

\maketitle

\begin{abstract}
We have succeeded in obtaining magnetized star models
that have extremely strong magnetic fields in the interior
of the stars. In our formulation,  arbitrary functions
of the magnetic flux function appear in the expression of
the current density. 
By appropriately choosing the functional form for
one of the arbitrary functions which corresponds
to the distribution of the {\it toroidal} current density, we have
obtained configurations with magnetic field distributions that
are highly localized within the central part and near the magnetic
axis region. The absolute values of the central magnetic fields are
stronger than those of the surface region by two orders of
magnitude. By applying our results to magnetars, 
the internal magnetic {\it poloidal} fields could be $10^{17}$ G, although the
surface magnetic fields are about $10^{15}$ G in the case of magnetars.
For white dwarfs, the internal magnetic {\it poloidal} fields could be $10^{12}$ G, when the
surface magnetic fields are $10^{9} - 10^{10}$ G .

\end{abstract}

\section{Introduction}

The magnetic field inside a star is scarcely detectable by direct
observations but has been considered to affect stellar evolutions and
activities in many aspects. For instance, if strong magnetic fields are 
hidden inside degenerate stars such as white dwarfs or neutron stars, 
they may significantly affect the cooling process of the stars
by providing an energy reservoir or by modifying heat conduction.  
Highly localized, anisotropic and relatively strong magnetic 
field configurations, on the other hand, may affect accretion 
modes onto degenerate stars in close binary systems
by providing a well-focused channel of accretion to their magnetic poles.
In order to know the possible distributions and strengths
of the magnetic fields inside the stars, we
have to rely on theoretical studies. Until very recently, however,
theoretical investigations could give us few hints about the interior
magnetic fields. The reason for that may be twofold: one is related to
the difficulty of the evolutionary computations of stellar magnetic fields
and the other is related to the lack of methods to obtain stationary
configurations of the magnetized stars.

Concerning the evolution of the stellar magnetic fields, since it has
been very difficult to pursue evolutionary computations of the global magnetic
fields for both interiors and exteriors of stars, few results have been
obtained. Recently, however, Braithwaite and his collaborators have succeeded
in following the evolution of global stellar magnetic fields
(\citealt{Braithwaite_Spruit_2004};
\citealt{Braithwaite_Nordlund_2006}; \citealt{Braithwaite_Spruit_2006};
\citealt{Braithwaite_2006}; \citealt{Braithwaite_2007};
\citealt{Braithwaite_2008}; \citealt{Braithwaite_2009}; \citealt{Duez_Braithwaite_Mathis_2010}).
 They found that the twisted-torus configurations of the magnetic fields inside
stars seem to be stable across the dynamical timescale.

On the other hand, to investigate possible structures of the interior and
exterior magnetic fields by imposing stationarity is
a different theoretical approach. 
Concerning this problem, many attempts have been made but it has 
also been difficult to obtain stellar structures with both {\it poloidal} and
{\it toroidal} non force-free magnetic fields self-consistently, not only in the Newtonian gravity but also in
general relativity (see e.g. \citealt{Chandrasekhar_Fermi_1953};
\citealt{Ferraro_1954}; \citealt{Chandrasekhar_1956};
\citealt{Chandrasekhar_Prendergast_1956}; \citealt{Prendergast_1956};
\citealt{Woltjer_1959a}; \citealt{Woltjer_1959b}; \citealt{Woltjer_1960};
\citealt{Wentzel_1961}; \citealt{Ostriker_Hartwick_1968};
\citealt{Miketinac_1973}; \citealt{Miketinac_1975};
\citealt{Bocquet_et_al_1995}; \citealt{Ioka_Sasaki_2004};
\citealt{Kiuchi_Yoshida_2008}; \citealt{Haskell_Samuelsson_Glampedakis_2008}
; \citealt{Duez_Mathis_2010}).
It is only recently that axisymmetric and stationary barotropic
stellar structures have been successfully solved for configurations with
both {\it poloidal} and {\it toroidal} magnetic components
(\citealt{Tomimura_Eriguchi_2005};
\citealt{Yoshida_Eriguchi_2006}; \citealt{Yoshida_Yoshida_Eriguchi_2006};
\citealt{Lander_Jones_2009}; \citealt{Otani_Takahashi_Eriguchi_2009})
in a non-perturbative manner.

It should be noted that the twisted-torus magnetic configuration that
appears during the evolutionary computations by \cite{Braithwaite_Spruit_2004}
is qualitatively the same as one of the exact axisymmetric and stationary
solutions obtained in \cite{Yoshida_Yoshida_Eriguchi_2006}. 
Moreover, stable configurations of stellar magnetic fields must have
a twisted-torus structure according to \cite{Braithwaite_2009}.
Concerning the stability analysis, this type of
configurations is expected to be stable, while magnetic
fields with purely {\it poloidal} configurations or purely
{\it toroidal} configurations have been shown to be unstable (see e.g.
\citealt{Tayler_1973_mnras}; \citealt{Wright_1973_mnras};
\citealt{Markey_Tayler_1973_mnras}; \citealt{Flowers_Ruderman_1977_apj}).

In this paper, we apply the formulation developed by \cite{Tomimura_Eriguchi_2005},
\cite{Yoshida_Eriguchi_2006} and \cite{Yoshida_Yoshida_Eriguchi_2006} in order 
to find out how strong and localized {\it poloidal} magnetic fields can 
exist inside stars,  as far as equilibrium configurations are 
concerned. In this formulation, the electric current density consists 
of several terms with different physical significances which contain 
arbitrary functionals of the magnetic flux function. 
These arbitrary functions correspond to the degrees 
of freedom in magnetized equilibria.
One of the arbitrary functionals in the expression for the 
electric current density corresponds to the current in the 
{\it toroidal} direction. 
By choosing this functional form properly, we would be able to 
obtain equilibrium configurations of axisymmetric barotropic stars 
with highly localized and extremely strong {\it poloidal}
magnetic fields.  

\section{Formulation and numerical method}

Since we employ the formulation developed by
\cite{Tomimura_Eriguchi_2005}, \cite{Yoshida_Eriguchi_2006},
and \cite{Yoshida_Yoshida_Eriguchi_2006}, here we summarize the
main scheme briefly and explain newly introduced parts in detail.

\subsection{Assumptions and basic equations}

We make the following assumptions for the magnetized stars.
\begin{enumerate}

  \item The system is in a stationary state, i.e.  $ \P{}{t} = 0$ \ .

  \item When stars are rotating and have magnetic fields, the rotational axis
      and the magnetic axis coincide.

 \item The rotation is rigid.

  \item The configurations are axisymmetric about the magnetic or
       the rotational axis, i.e.
       $ \P{}{\varphi} = 0 $, where we use the spherical
       coordinates $(r, \theta, \varphi)$.

  \item The configurations are symmetric with respect to the equator.

 \item There are no meridional flows.

 \item The star is self-gravitating.

 \item The systems are treated in the framework of non-relativistic physics.
  
  \item The conductivity of the stellar matter is 
   infinite, i.e. the ideal magnetohydrodynamics (MHD) 
	approximation is employed.

  \item No electric current is assumed in the vacuum region.

  \item The barotropic  equation of state is assumed :
\begin{eqnarray}
   p = p(\rho) \ .
\end{eqnarray}
\end{enumerate}
Here $p$ and $\rho$ are the pressure and the mass density, respectively.
Assumptions of axisymmetric and equatorial symmetries 
as well as rigid rotation are adopted here 
in order to simplify our investigations.

In a rotating star under a radiative equilibrium,  there appears to be
 meridional flow in special cases. However, we neglect it because the time scale
is many orders of magnitude larger than the (magneto)hydrodynamic one
(\citealt{Tassoul_2000}). Also, there is a suggestion that gradual diffusion of
the internal magnetic fields drives a meridional flow (\citealt{Urpin_Ray_1994}). 
The time scale of this, again, 
is much larger than the (magneto)hydrodynamic one. Thus it is
also neglected.

Under these assumptions, the basic equations are written as follows.
The continuity equation is expressed as
\begin{eqnarray}
   \nabla \cdot (\rho \Vec{v}) = 0 \ ,
\end{eqnarray}
where $\Vec{v}$ is the fluid velocity. The equations of motion in the
stationary state are written as:
\begin{eqnarray}
   \frac{1}{\rho} \nabla p = -\nabla \phi_g +
   R \Omega^2 \Vec{e}_R+ \frac{1}{\rho} \left( \frac{\Vec{j}}{c} \times
   \Vec{H} \right),
   \label{Eq:eular}
\end{eqnarray}
where $\phi_g$, $\Omega$, $\Vec{j}$, $c$ and $\Vec{H}$ are 
gravitational potential, angular velocity, electric current density,
speed of light and magnetic field, respectively. Here we 
use the cylindrical coordinates $(R, \varphi, z)$ and $\Vec{e}_R$
is the unit vector in the $R$-direction. The gravitational potential
satisfies Poisson equation:
\begin{eqnarray}
   \Delta \phi_g = 4 \pi G \rho \ ,
   \label{Eq:Poisson}
\end{eqnarray}
where $G$ is the gravitational constant. Maxwell's
equations are written as,
\begin{eqnarray}
   \nabla \cdot \Vec{E} = 4\pi \rho_e \ ,
\end{eqnarray}
\begin{eqnarray}
   \nabla \cdot \Vec{H} = 0 \ ,
   \label{Eq:divH}
\end{eqnarray}
\begin{eqnarray}
   \nabla \times \Vec{E} = 0 \ ,
   \label{Eq:rotE}
\end{eqnarray}
\begin{eqnarray}
   \nabla \times \Vec{H} = 4 \pi \frac{\Vec{j}}{c} \ ,
   \label{Eq:rotH}
\end{eqnarray}
where $\rho_e$ and $\Vec{E}$ are the electric charge density and
the electric field, respectively. 
Notice that we neglect the displacement current term in Eq.(\ref{Eq:rotE})
 as is common in MHD approximation.
The ideal MHD condition, or the
generalized Ohm's equation, can be expressed as:
\begin{eqnarray}
   \Vec{E} = - \frac{\Vec{v}}{c} \times \Vec{H} \ .
\end{eqnarray}
We choose two kinds of barotropic equations of state.
One is the polytropic equation of state:
\begin{eqnarray}
   p = K_0 \rho^{1+1/N} \ ,
\end{eqnarray}
where $N$ and $K_0$ are the polytropic index and the polytropic constant,
respectively. 
The other is the degenerated Fermi gas at  zero temperature,
defined as
 \begin{eqnarray}
p = a[x(2x^2 - 3)\sqrt{x^2 + 1} + 3 \ln(x + \sqrt{x^2 + 1}) ],
\end{eqnarray}
where
\begin{eqnarray}
 \rho = bx,
\end{eqnarray}
\begin{eqnarray}
 a = 6.00 \times 10^{22} \hspace{10pt} \mathrm{dyn/cm}^2,
\end{eqnarray}
\begin{eqnarray}
 b = 9.825 \times 10^5  \mu_e \hspace{10pt} \mathrm{g/cm}^3.
\end{eqnarray}
Here $\mu_e$ is the mean molecular weight.
We fix $\mu_e=2$ in all our computations here,
which corresponds to a fully ionized pure hydrogen gas.
This choice of parameters is same as that in \cite{Hachisu_1986}.

\subsection{The form of the current density and the boundary condition}
\label{sec:bc}

From the assumptions of axisymmetry and stationarity,
we introduce magnetic flux function $\Psi$ as follows:
\begin{eqnarray}
   H_R \equiv -\frac{1}{R}\P{\Psi}{z} , \hspace{10pt}
   H_z \equiv \frac{1}{R}\P{\Psi}{R} \ ,
\end{eqnarray}
where $H_R$ and $H_z$ are magnetic field components in the $R$-direction
and $z$-direction, respectively. We assume this flux function is positive 
in the entire space.
By introducing this magnetic flux function,
equation (\ref{Eq:divH}) can be automatically satisfied. It should be noted
that the magnetic flux function $\Psi$ can be expressed as:
\begin{eqnarray}
    \Psi = r \sin \theta A_{\varphi} \ , 
\end{eqnarray}
where $A_{\varphi}$ is the $\varphi$-component of the
vector potential $\Vec{A} = (A_R, A_{\varphi}, A_z)$. 

As shown in \cite{Tomimura_Eriguchi_2005}, for axisymmetric and
stationary barotropes with rigid rotation we can constrain the form
of the electric current density by using an integrability 
condition of the equations of motion, equation (\ref{Eq:eular}):
\begin{eqnarray}
  \frac{\Vec{j}}{c} = \frac{1}{4\pi} \D{\kappa(\Psi)}{\Psi} \Vec{H} + r \sin \theta \rho 
   \mu(\Psi) \Vec{e}_\varphi \ ,
   \label{Eq:current}
\end{eqnarray}
where $\kappa(\Psi)$ and $\mu(\Psi)$  are arbitrary functions of
the magnetic flux function $\Psi$. 
Notice, in particular, that the 
{\it toroidal} component of magnetic field is given as
\begin{equation}
H_\varphi = \frac{\kappa (\Psi)}{r \sin \theta}
\label{Eq:H_phi}
\end{equation}
which can be derived from equations (\ref{Eq:eular}),  (\ref{Eq:rotH})
 and (\ref{Eq:current}). 
It should be noted that these two arbitrary functions 
are conserved along the {\it poloidal} magnetic field lines.
Although the meanings of these two functions are described in 
previous works (see, e.g.,  \citealt{Lovelace_et_al_1986}),
in this paper we will explain their meanings differently 
from our point of view. 

Since we have assumed that there is no electric current 
in the vacuum region, in other words that there is no 
{\it toroidal} magnetic field outside the star 
(see equation \ref{Eq:current}), the form for $\kappa$ 
needs to be a special one. The simplest form can be
$\kappa=$ constant (\citealt{Ioka_Sasaki_2004}, 
\citealt{Haskell_Samuelsson_Glampedakis_2008}), 
but for this choice of $\kappa$ the {\it toroidal} magnetic 
field would extend to the vacuum region. In order to 
avoid this possibility, we choose the functional form 
of $\kappa$ as follows:
\begin{eqnarray}
  \kappa (\Psi) =
   \left\{
     \begin{array}{lr}
        0 \ , & \mathrm{for} \hspace{10pt} \Psi \leq \Psi_{\max} \ , \\
        \dfrac{\kappa_0}{k+1}(\Psi - \Psi_{\max})^{k+1} \ , 
         & \mathrm{for} \hspace{10pt} \Psi \geq \Psi_{\max} \ ,
     \end{array}
   \right.
   \label{Eq:kappa}
\end{eqnarray}
This choice of $\kappa$ is the same as that in 
\cite{Yoshida_Eriguchi_2006} and 
\cite{Lander_Jones_2009}. In this paper we 
fix $k=0.1$. 
Equations (\ref{Eq:H_phi}) and (\ref{Eq:kappa}) ensure that the 
{\it toroidal} magnetic field  vanishes smoothly at the stellar surface. 
Incidentally, using these functionals,
we obtain the first integral of equation (\ref{Eq:eular}) 
as follows:
\begin{eqnarray}
    \int \frac{dp}{\rho} = -\phi_g + \frac{1}{2}(r \sin \theta)^2\Omega^2_0
   + \int \mu(\Psi) \, d\Psi + C \ ,
   \label{Eq:first_int}
\end{eqnarray}
where $C$ is an integration constant. The first term of the right-hand side is the 
gravitational potential. The second term on the right hand side 
is related to rotation. We can consider it as a rotational potential.
Similarly, the third term means the potential of Lorentz force. 
We can regard this term as the magnetic force potential.
Therefore, $\int \mu \, d\Psi$ is considered to be non-force-free 
contribution from the current density, as is seen in equation (\ref{Eq:current}).
Since the Lorentz force is given by the cross product $\Vec{j} / c \times \Vec{H}$,
the first term of equation (\ref{Eq:current}) 
has no effect on the equation of motion, i.e, it is force-free,
and only the second term contributes to the Lorentz force, 
i.e., non force-free.
The distribution of Lorentz force could be changed by adopting different 
functional forms for $\mu$. All previous works 
(\citealt{Tomimura_Eriguchi_2005}, \citealt{Yoshida_Eriguchi_2006},
\citealt{Yoshida_Yoshida_Eriguchi_2006}, \citealt{Lander_Jones_2009}, 
\citealt{Otani_Takahashi_Eriguchi_2009}) fixed $\mu = \mu_0 $ (constant).
We choose a different functional form for $\mu$ in this paper as follows:
\begin{eqnarray}
   \mu(\Psi)                & = & \mu_0 (\Psi + \epsilon )^m \ , \\
   \int \mu(\Psi) \, d \Psi & = & \frac{\mu_0}{m+1}(\Psi + \epsilon )^{m+1} \ ,
   \label{eq:muuu}
\end{eqnarray}
where $m$ and $\epsilon$ are two constant parameters. 
In order to avoid singular behavior,  we fix 
$\epsilon = 1.0 \times 10^{-6}$ in all calculations.
As we shall see below, the parameter $m$ determines a degree of 
localization of the interior {\it poloidal} magnetic field. 
We assume that {\it poloidal} magnetic fields extend throughout 
the whole space and that there are no discontinuities even at the 
stellar surface. The global magnetic field configurations of our models 
are nearly dipole-like because of the requirement of the
functional form for $\kappa$  at the stellar surface.
These configurations contain closed {\it poloidal} magnetic field lines 
inside the star. The flux function $\Psi$ attains its maximum 
at the central parts of these closed field lines and it takes its
minimum on the symmetric axis and at infinity. The minimum value is 
zero because of  $\Psi = r \sin \theta A_\varphi$ and 
the boundary condition for $A_\varphi = 0 $ at infinity. 
The magnetic potential $\left(\int\mu d\Psi\right)$ changes its
qualitative behavior in its spatial distribution when $m=-1$.
If we adopt $m < -1$, as $\Psi$ decreases from its maximum to 
zero on the axis of the star
the value of the magnetic potential increases
unboundedly if $\epsilon\to 0$.
As a result, the {\it poloidal} magnetic field lines are concentrated 
near the axis in order to fulfill such magnetic potential 
distributions. On the other hand, if we choose $m > - 1$, 
the value of the magnetic 
potential decreases as $\Psi$ decreases from its maximum 
to zero, which is realized on the axis.
Then the {\it poloidal} magnetic field lines are
distributed more uniformly than those for configurations with
$m < -1$.  If we choose $m=0$, we obtain $\mu = $ constant 
configurations. 
They are the same as those investigated by
other authors. 

It is remarkable that the only freedom that
we can take in our formulation is related to the choices of
functional forms and the values of the parameters which
appear in those functions. It implies that degrees of freedom 
for choices for these functions and parameters correspond 
to degrees of freedom for many kinds of stationary 
axisymmetric magnetic field configurations. In fact, as we 
see from our results, different values for $m$ result
in qualitatively different distributions for the  magnetic 
potentials and the {\it poloidal} magnetic fields. In other words, 
we can control the magnetic field distributions to a certain extent 
by adjusting the value for $m$. This is the reason why 
we use this functional form of $\mu$ in this paper.

After we choose the functional form of the current density,
by using Eq.(\ref{Eq:current}) and the definition of the vector
potential, we obtain the following partial differential equation 
of the elliptic type:
\begin{eqnarray}
   \Delta (A_\varphi \sin\varphi) 
    = 4 \pi S_A(r,\theta) \sin\varphi \ , \hspace{10pt}
    S_A \equiv - \frac{j_\varphi}{c}.
   \label{Eq:Poisson_A}
\end{eqnarray}
As we have seen in the previous paragraph, all the physical 
quantities related to the vector potential can be expressed solely by 
$\Psi$. Therefore we need not solve for $A_R$ and $A_z$. 
It implies that our present formulation does not depend on 
the gauge condition for the vector potential $\Vec{A}$. 
Next we impose the boundary conditions for the gravitational 
potential and the vector potential, chosen as follows:
\begin{eqnarray}
   \phi_g \sim {\cal O} \left(\frac{1}{r}\right) \ ,
   \hspace{10pt}(r \rightarrow \infty) \ ,
   \label{Eq:BC_gphi}
\end{eqnarray}
\begin{eqnarray}
   A_\varphi \sim  {\cal O} \left(\frac{1}{r}\right) \ ,
   \hspace{10pt}(r \rightarrow \infty) \ .
   \label{Eq:BC_A}
\end{eqnarray}
This boundary condition for $A_\varphi$ results in
\begin{eqnarray}
   H_p \sim {\cal O} \left(\frac{1}{r^2}\right) \ ,
   \hspace{10pt}(r \rightarrow \infty) \ .
   \label{Eq:BC_H}
\end{eqnarray}
where $H_p$ is the {\it poloidal} magnetic field.
From these boundary conditions and using a proper Green's function
for the Laplacian, we have the integral representations of
equation (\ref{Eq:Poisson}) and equation (\ref{Eq:Poisson_A}) as follows:
\begin{equation}
   \phi_g(\Vec{r}) = -G \int \frac{\rho(\Vec{r}')}{|\Vec{r} - \Vec{r}'|}
   \, d^3 \Vec{r}' \ ,
\end{equation}
\begin{equation}
   A_\varphi(\Vec{r}) \sin \varphi
   = - \int \frac{S_A(\Vec{r}')\sin\varphi'}
   {|\Vec{r} - \Vec{r}'|} \, d^3 \Vec{r}' \ .
\end{equation}
Therefore, we can obtain smooth potentials, $\phi_g$ and $A_\varphi$ 
by integrating these equations. 
Since we have chosen the functional form of the current density which 
decreases near the surface and vanishes at the stellar surface sufficiently smoothly, 
we obtain continuous {\it poloidal} magnetic fields from $A_\varphi$.
%

\subsection{Global characteristics of equilibria}

To see the global characteristic of magnetized equilibria,
we define some integrated quantities as follows:
\begin{eqnarray}
   W \equiv \frac{1}{2}\int \phi_g \rho \, d^3 \Vec{r} \ ,
\end{eqnarray}
\begin{eqnarray}
   T \equiv \frac{1}{2} \int \rho (R \Omega)^2 \, d^3 \Vec{r} \ ,
\end{eqnarray}
\begin{eqnarray}
   \Pi \equiv \int p \, d^3 \Vec{r} \ ,
\end{eqnarray}
\begin{eqnarray}
   U \equiv N \Pi \ ,
\end{eqnarray}
for polytropic models and
\begin{eqnarray}
   U \equiv \int g(x) \, d^3 \Vec{r} \ ,
\end{eqnarray}
\begin{eqnarray}
 g(x) = a\{ 8x^3 [(x^2 + 1)^\frac{1}{2} - 1]\} - p, 
\end{eqnarray}
for the Fermi gas configurations (see. 
\citealt{Chandrasekhar_stellar_structure}),
\begin{eqnarray}
   {\cal H} \equiv \int r \cdot \left(\frac{\Vec{j}}{c} \times \Vec{H} \right) \, d^3 \Vec{r} \ .
\end{eqnarray}
\begin{eqnarray}
   K = \int (\nabla \times \Vec{A}) \cdot \Vec{A} \, d^3 \Vec{r}
   = \int \Vec{H} \cdot \Vec{A} \, d^3 \Vec{r} \ ,
\end{eqnarray}
where $W$, $T$, $\Pi$, $U$, ${\cal H}$ and $K$ are the 
gravitational energy, rotational energy, total pressure, 
internal energy, magnetic field energy
and magnetic helicity, respectively. 
In order to evaluate the structures of magnetic fields, 
we define some physical quantities 
related to the magnetic fields as follows:
\begin{eqnarray}
   H_{sur} = \dfrac{\int_0^{2\pi} 
   \int_0^{\pi} \, 
 r_s^2(\theta) \, 
\sin \theta 
   |\Vec{H}(r_s, \theta)|
 \, d\theta
d\varphi
}{S} \ ,
\end{eqnarray}
where $r_s(\theta)$ and $|H_{sur}|$ are the stellar
radius in the direction of $\theta$ and the surface magnetic 
field strength, respectively, and
the surface area of the star is defined as:
\begin{eqnarray}
  S = \int_0^{2\pi} 
\int_0^{\pi}    \,  r_s^2(\theta) \sin \theta \, d\theta 
\, d \varphi .
\end{eqnarray}
The volume-averaged magnetic field strength in the central region
of the star is defined as
\begin{eqnarray}
   H_\mathrm{c} = 
   \dfrac{\int_0^{2\pi}
   \int_0^\pi \,
   \int_0^{r_c} \,  
r^2 \sin \theta |\Vec{H}(r,\theta)| dr d\theta d\varphi}
   {V},
   \label{Eq:H_c}
\end{eqnarray}
where we choose  $r_c = 0.01 r_e$ and $V$ is the volume of the
central region with $r \le r_c$, defined as
\begin{eqnarray}
   V = \int_0^{2\pi} \, 
\int_0^\pi 
\int_0^{r_c} r^2\, 
\sin \theta dr 
\, d\theta
d\varphi \ .
\end{eqnarray}
This central region seems to be very small, but we can resolve
it sufficiently because we use non-uniform and centrally concentrated 
meshes (see. Fig. \ref{fig:mesh} and  Eq. \ref{Eq:r_mesh} in Appendix).
We have $77$ meshes to resolve the region in actual numerical computations.

In order to know the contributions of the {\it poloidal} magnetic field and
the {\it toroidal} magnetic field separately, we define the {\it poloidal}
magnetic energy ${\cal H}_p$ and the {\it toroidal} magnetic energy ${\cal H}_t$ as
\begin{eqnarray}
   {\cal H}_p = \frac{1}{8\pi}
    \int_0^{2 \pi} \int_0^{\pi} \int_0^{\infty}
r^2 \, \sin \theta \,  
 |H_r(r,\theta)^2 + H_\theta(r,\theta)^2| dr d\theta d\varphi \ ,
\end{eqnarray}
\begin{eqnarray}
   {\cal H}_t = \frac{1}{8\pi}
    \int_0^{2 \pi} \int_0^{\pi} \int_0^{\infty}
r^2 \, \sin \theta \,  
 |H_\varphi(r,\theta)^2| dr d\theta d\varphi \ ,
\end{eqnarray}
As for the magnetic multipole moment seen outside a star,
we compute each multipole component by
solving the following equation in a vacuum:
\begin{eqnarray}
   \Delta \left(A_\varphi  \sin \varphi\right) = 0 \ .
\end{eqnarray}
Considering the boundary conditions at infinity and the symmetry
of the magnetized stars, the solution of the above equation can be
expressed as
\begin{eqnarray}
   A_\varphi \sin \varphi
             \equiv  \sum_{n=1}^{\infty} A_{\varphi, n} \sin \varphi
             = \sum_{n=1}^{\infty} b_{n,1} r^{-n-1} Y_{n,1}(\theta,\varphi) \ ,
   \label{Eq:bn}
\end{eqnarray}
where $Y_{n,1}(\theta,\varphi)$ is the spherical harmonics of degree $n$ and order $m = 1$.
The coefficients $b_{n,1}$ correspond to the magnetic multipoles.
%

\subsection{Setting for Numerical Computations}

For numerical computations, the physical quantities are transformed
into dimensionless ones using the maximum density $\rho_{\max}$,
the maximum pressure $p_{\max}$ and the equatorial radius $r_e$ as 
follows:
\begin{eqnarray}
   \hat{r} \equiv \frac{r}{r_e} = \frac{r}{
   \sqrt{\frac{1}{\alpha}\frac{p_{\mathrm{max}}}{4\pi G
   \rho_{\mathrm{max}}^2}}} \ ,
\end{eqnarray}
for polytropic configurations and
\begin{eqnarray}
   \hat{r} \equiv \frac{r}{r_e} = \frac{r}{
   \sqrt{\frac{1}{\alpha}\frac{8a}{b}  \frac{1}{4\pi G
   \rho_{\mathrm{max}}^2}}} \ ,
\end{eqnarray}
for the Fermi gas models, and 
\begin{eqnarray}
   \hat{\rho} \equiv \frac{\rho}{\rho_\mathrm{max}} \ .
\end{eqnarray}
Here $\alpha$ is introduced so as to make the distance from the center
to the equatorial surface of the star to be unity.
Arbitrary functions are also transformed into dimensionless ones.
Quantities with $\hat{}$ are dimensionless. 
For example, the dimensionless length is $\hat{r}$
and the dimensionless arbitrary functions are 
$\hat{\mu}$ and  $\hat{\kappa}$, respectively. 
Dimensionless forms of other quantities are collected
in Appendix \ref{App:dimensionless}.

The computational domain is defined as $0\leq\theta\leq\frac{\pi}{2}$ in the angular
direction and $0\leq \hat{r}\leq 2$ in the radial direction. Since the 
equation of magnetohydrostationary equilibrium is defined 
only inside the star and the source terms of the elliptic equations for the gravitational 
potential and the magnetic flux function vanish outside the star, 
our computational domain covers a region of the space
that is sufficient for obtaining equilibria.
In order to resolve the region near the axis sufficiently, 
we use a special coordinate in actual numerical computations.
Total mesh numbers in $r$-direction and in $\theta$-direction
are 1025 and 1025, respectively. 
We describe details of the computational grid points 
in Appendix \ref{App:grid}.

\subsection{Numerical method}

We use the scheme of \cite{Tomimura_Eriguchi_2005}.
This scheme is based on the Hachisu Self-Consistent Field (HSCF) scheme 
(\citealt{Hachisu_1986}),  which is the method 
for obtaining equilibrium configurations of
rotating stars. We define the ratio of the equatorial radius to
the polar radius as the axis ratio $q$. This quantity $q$ characterizes 
how distorted the stars are due to non-spherical forces. 
The stronger the non-spherical force becomes, the more 
distorted the stellar shape is. The non-spherical force can be
the centrifugal force, the magnetic force or both of them.
We fix the value of $q$ in order to obtain the magnetized 
equilibria. We also fix one of $\hat{\mu}_0$ and $\hat{\Omega}_0$.
If we fix $\hat{\mu}_0$, we will obtain the value of $\hat{\Omega}_0$
after the  relaxation and iteration. If we fix  $\hat{\Omega}_0$, 
we will obtain $\hat{\mu}_0$. Then, we will obtain one 
magnetized equilibrium state.

\subsection{Numerical accuracy check}

\begin{figure}
 \begin{center}
  \includegraphics[scale=0.6]{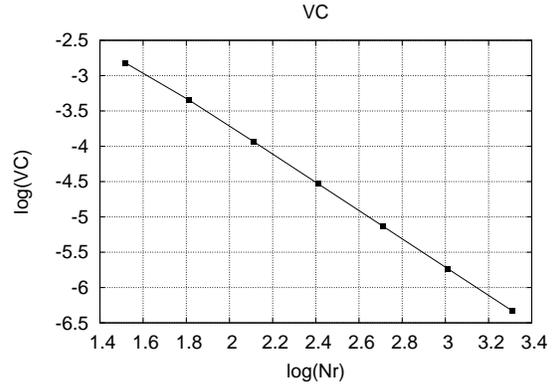}
   \caption{The virial quantity VC, plotted against
            the number of grid points in the $r$-direction.}
   \label{fig:VC}
 \end{center}
\end{figure}

In order to check the accuracy of converged solutions,  we
compute a relative value of the virial relation
as follows:
\begin{eqnarray}
   \mathrm{VC} \equiv \frac{|2T + W + 3\Pi +{\cal H}|}{|W|}.
\end{eqnarray}
Since this quantity VC must vanish for exact equilibrium
configurations, we can check the global accuracies of the numerically
obtained models as a whole (see e.g. \citealt{Hachisu_1986}).
Since the numerical results depend on mesh size, we have
computed the same model by changing the number of grid points
in the $r$-coordinate but fixing the number of grid points
in the $\theta$-direction as $n_\theta = 513$. 
Fig. \ref{fig:VC} shows VC as a function of the number of
grid points in the $r$-coordinate for polytropic models.
Since we use schemes of second-order accuracy, 
VC decreases as the square inverse of the number of grid 
points (see also \citealt{Lander_Jones_2009}; 
\citealt{Otani_Takahashi_Eriguchi_2009}).

\section{Numerical Results}

We give a brief summary of our numerical results here.
First we show the basic features for negative $m$ models 
and the dependences  of the magnetic field 
configurations on the values of $m$ for barotropes.
We also show rotating and magnetized polytropic models
in order to examine the effect of rotation on  
magnetic fields.  The influence of
the equation of state on the interior magnetic field
is also displayed.
We have computed $N=0.5, 1, 1.5$ polytropic models
and four white dwarf models  with 
$\rho_c = 1.0\times 10^7, 1.0 \times 10^8,$ 
$1.0 \times 10^9,$ and $1.0 \times 10^{10} \mathrm{g cm^{-3}}$.

\subsection{Effect of the distribution of the {\it toroidal} 
current density on the distribution of the magnetic field}
\label{Sec:current}

\begin{table*}
\begin{minipage}{150mm}
\begin{tabular}{cccccccccccccccc}
\hline
$m$ &  $1-q$ &   $H_\mathrm{c} / H_\mathrm{sur}$ &
${\cal H}_p/{\cal H}$ &  ${\cal H}/|W|$ & $\Pi/|W|$ & $\alpha$ & $\hat{\mu}_0$ &
$\hat{K}$  &VC\\
\hline
&&&&& \hspace{-80pt} $N = 1.0$ &  \\
\hline
 -2.0 &  2.2E-2  &  1.03E+2 &9.987E-1 &  3.74E-5 &
  3.33E-1 &  5.07E-2 &  2.28E-9 &  7.76E-7 &  5.132E-8 \\

-1.5 &  1.9E-3 &  4.44E+1 &9.982E-1 &  3.02E-5 &
  3.33E-1 &  5.07E-2 &  9.65E-8 &  8.27E-7 & 2.646E-6 \\
 
-1.1 &  4.2E-4 &  2.19E+1 &9.978E-1 &  2.60E-5 &
  3.33E-1 &  5.07E-2 &  1.80E-6 &  8.33E-7 & 2.775E-6 \\
 
-0.9 & 2.5E-4 &  1.62E+1 &9.976E-1 &  2.45E-5 &
  3.33E-1 &  5.07E-2 &  7.74E-6 &  8.35E-7 & 2.785E-6 \\
 
-0.5 & 1.3E-4  & 1.02E+1 &9.972E-1 &  2.21E-5 &
  3.33E-1 &  5.07E-2 &  1.41E-4 &  8.35E-7 & 2.788E-6 \\
 
0.0 &  8.8E-5 &  7.17E+0 &9.968E-1 &  1.99E-5 &
  3.33E-1 &  5.07E-2 &  5.25E-3 &  8.31E-7 & 2.789E-6 \\
 
0.5 &  7.0E-5 & 5.69E+0 &9.963E-1 &  1.83E-5 &
  3.33E-1 &  5.07E-2 &  1.92E-1 &  8.23E-7 & 2.789E-6 \\
 
1.0 &  6.1E-5&  4.78E+0 &9.959E-1 &  1.70E-5 &
  3.33E-1 &   5.07E-2 &  6.90E+0 & 8.11E-7 & 2.789E-6 \\

 \hline
\end{tabular}

 \caption{Physical quantities for $ \hat{\Omega}_0 = 0$, 
$\hat{\kappa}_0 = 10$ and $H_{sur} = 1.5\times10^{15} 
\mathrm{G }$ polytropes with different values of $m$.}

\label{tab:m-mu}
\end{minipage}
\end{table*}

\begin{figure*}
\begin{minipage}{150mm}

\includegraphics[scale=0.75]{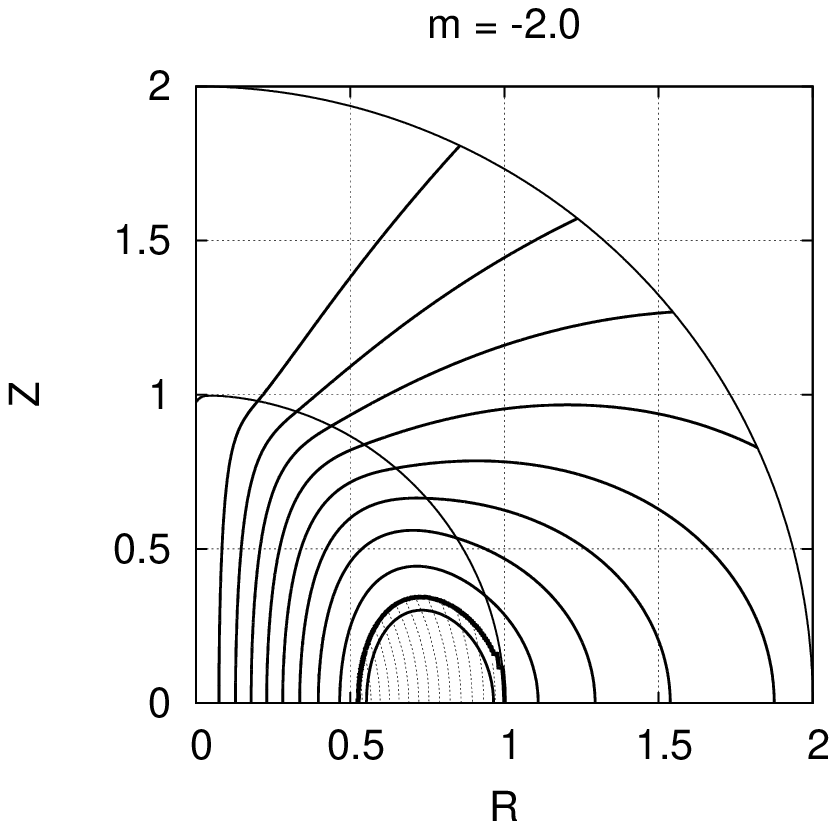}
\includegraphics[scale=0.75]{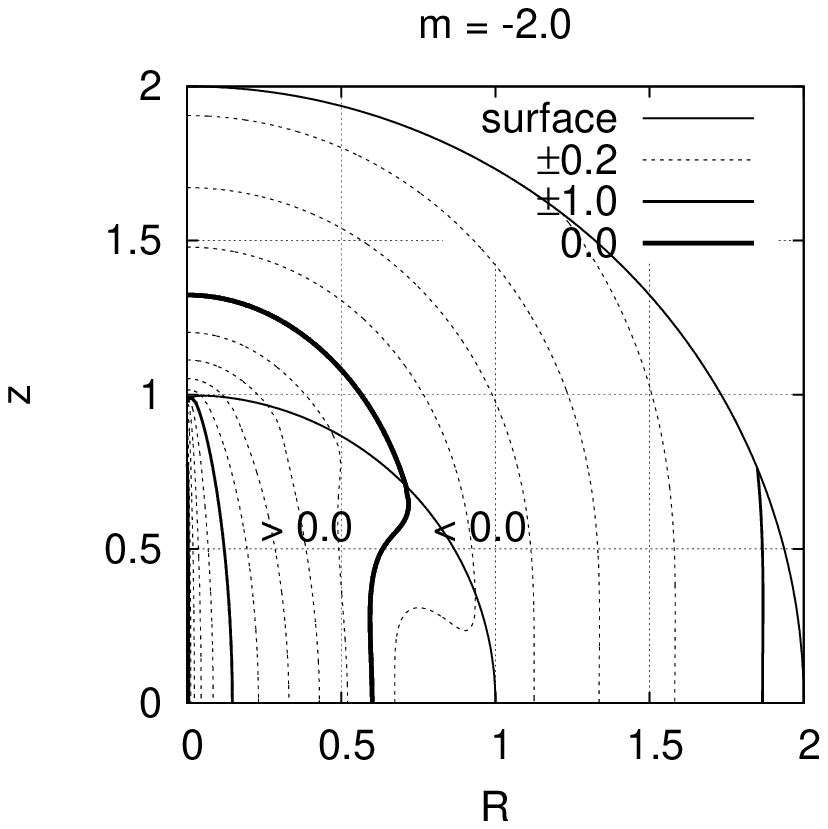}

\includegraphics[scale=0.75]{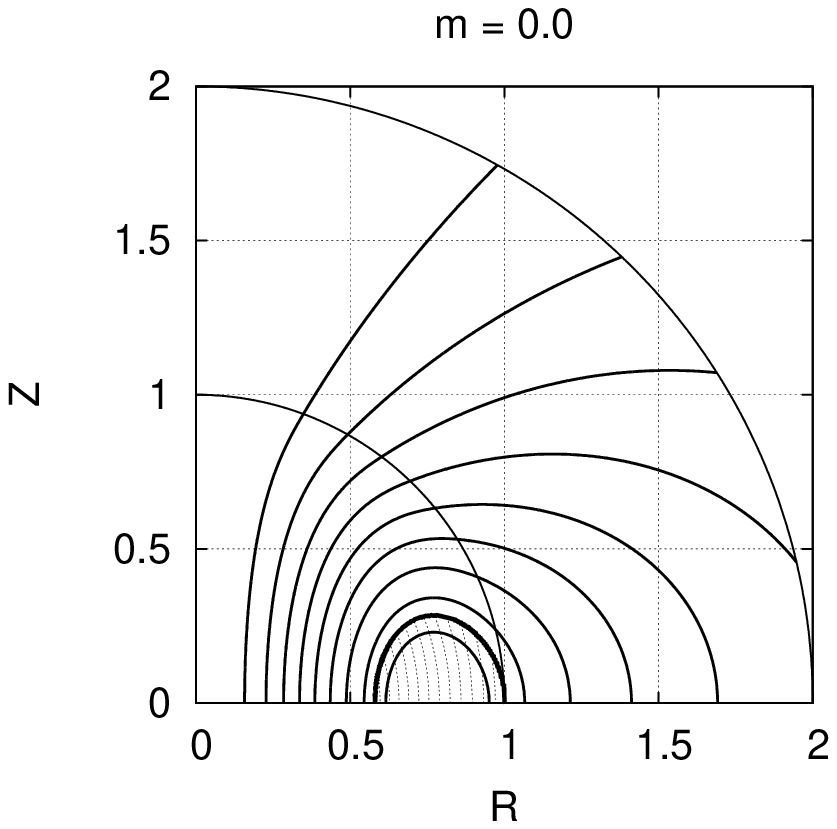}
\includegraphics[scale=0.75]{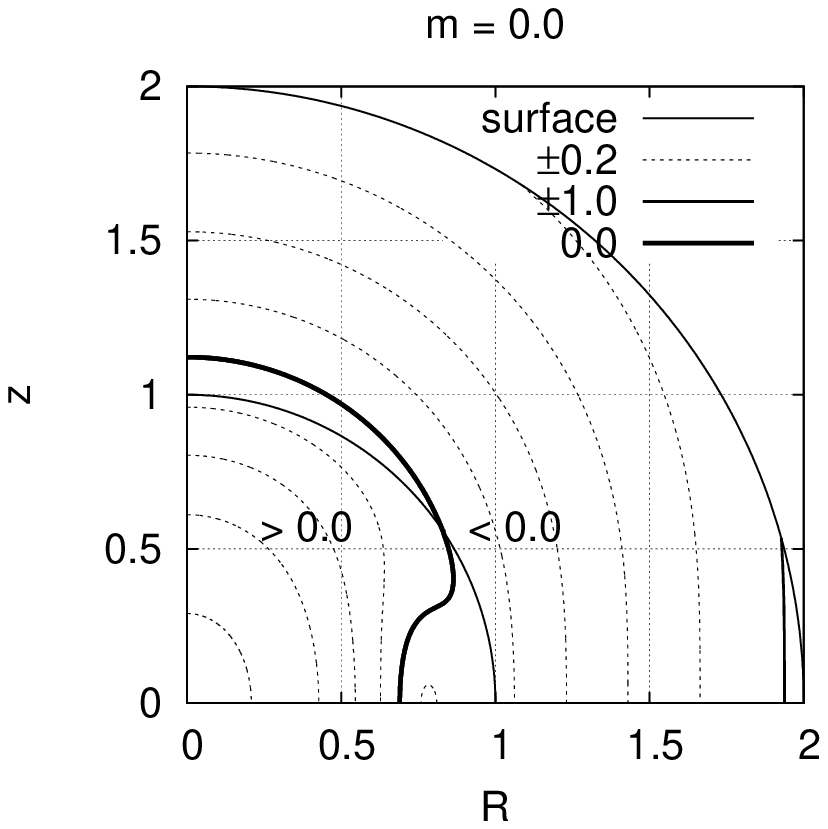}

\includegraphics[scale=0.75]{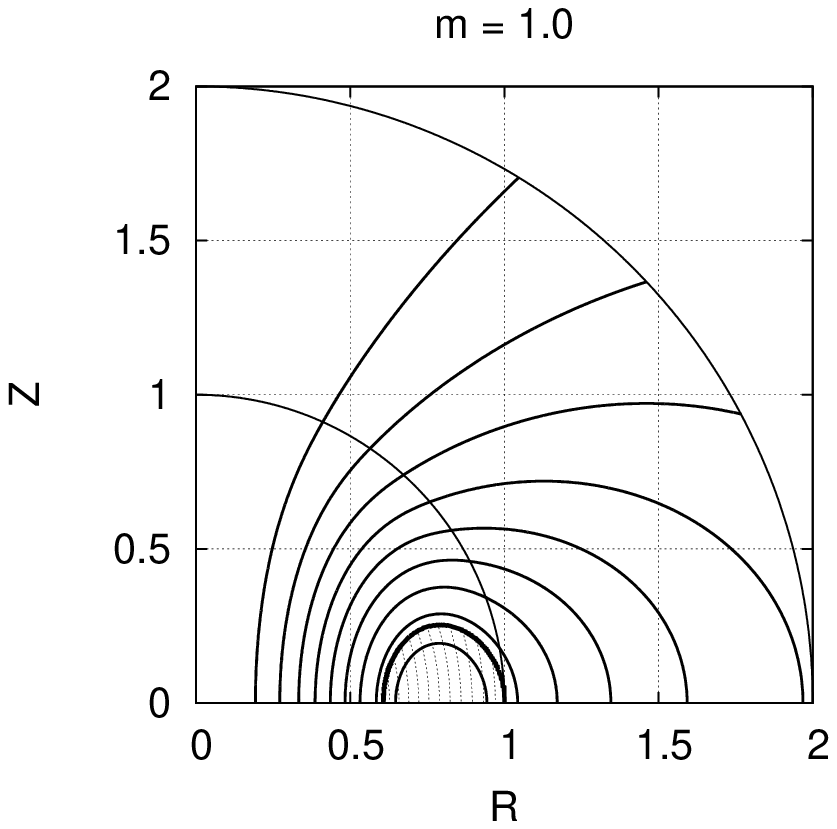}
\includegraphics[scale=0.75]{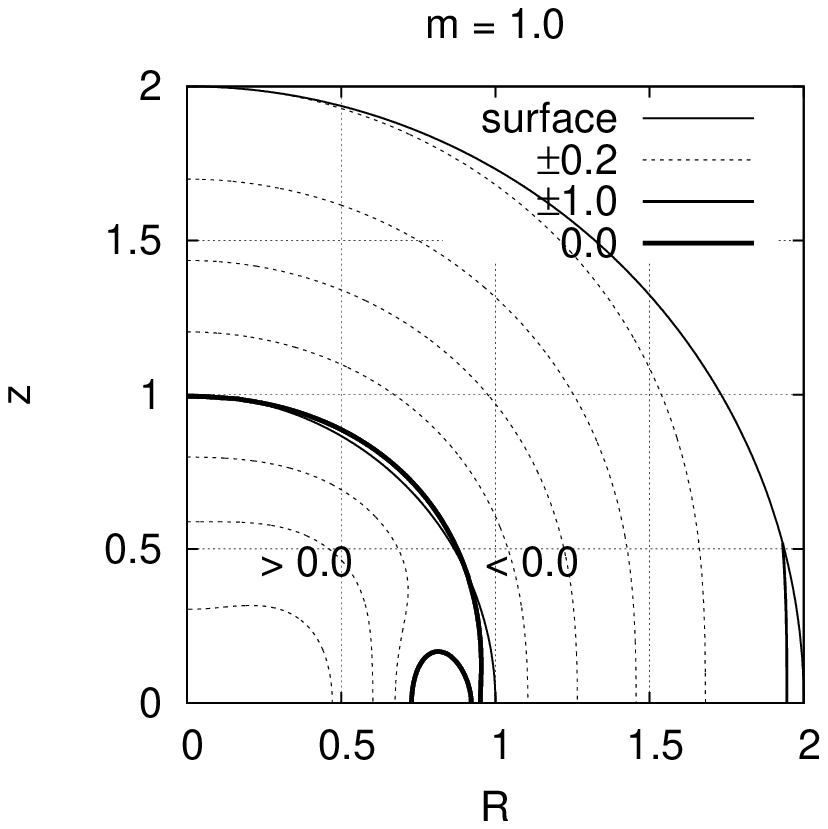}
 
 \caption{Contours for the magnetic flux function (left panels)
and for the logarithm of the strength of the magnetic field normalized by the
averaged surface magnetic field (right panels) are shown.
The inner solid circle corresponds to the surface of the star and 
the outer solid circle denotes the boundary 
of our computational region. 
The flux difference between two adjacent contours of thick lines is 
1/10 of the maximum value of $\hat{\Psi}$.
In the left panels, the thick {\it poloidal} field line is the boundary 
of the {\it toroidal} magnetic field region. 
The {\it toroidal} magnetic field exists only inside the region.
In the right panels, the distribution of the logarithm of the 
magnetic field normalized by the averaged surface magnetic field, 
$\log_{10}|\Vec{H}/H_{sur}|$ contour, is shown. The thick solid 
curve corresponds to the curve with 
$\log_{10} |\Vec{H}/ H_{sur}| = 0$.  Inside this curve
$\log_{10} |\Vec{H} / H_{sur}| > 0$ and outside this curve
$\log_{10} |\Vec{H} / H_{sur}| < 0$. The difference between 
two adjacent contours is 0.2. 
}
\label{fig:magnetic_structure}
\end{minipage}
\end{figure*}

 \begin{figure*} 
  \begin{minipage}{150mm}
  \begin{center}
  \includegraphics[scale=0.75]{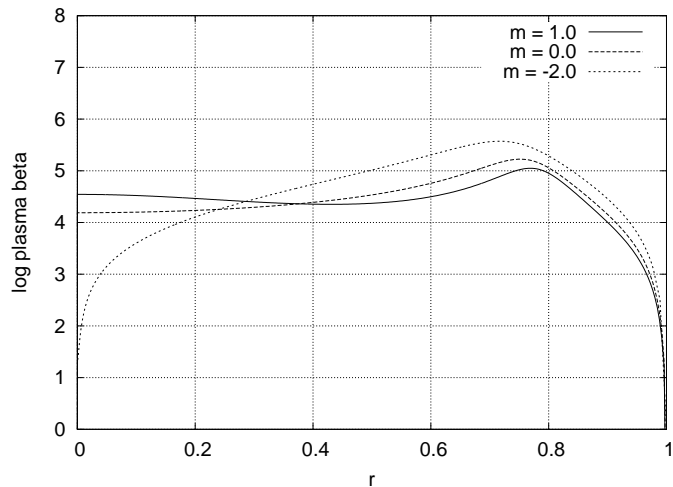}
  \caption{Profiles of $\log \beta$ at $\theta = \pi/2$,
  where $\beta$ is the plasma $\beta$. Solid line represents
  the distribution for an $m=1.0$ configuration, dashed line that
  for  an $m=0.0$ configuration and dotted line that for an 
  $m=-2.0$ model, respectively.}
  \label{fig:beta}
  \end{center}
  \end{minipage}
\end{figure*}

\begin{figure*}
\begin{minipage}{150mm}
\includegraphics[scale=0.55]{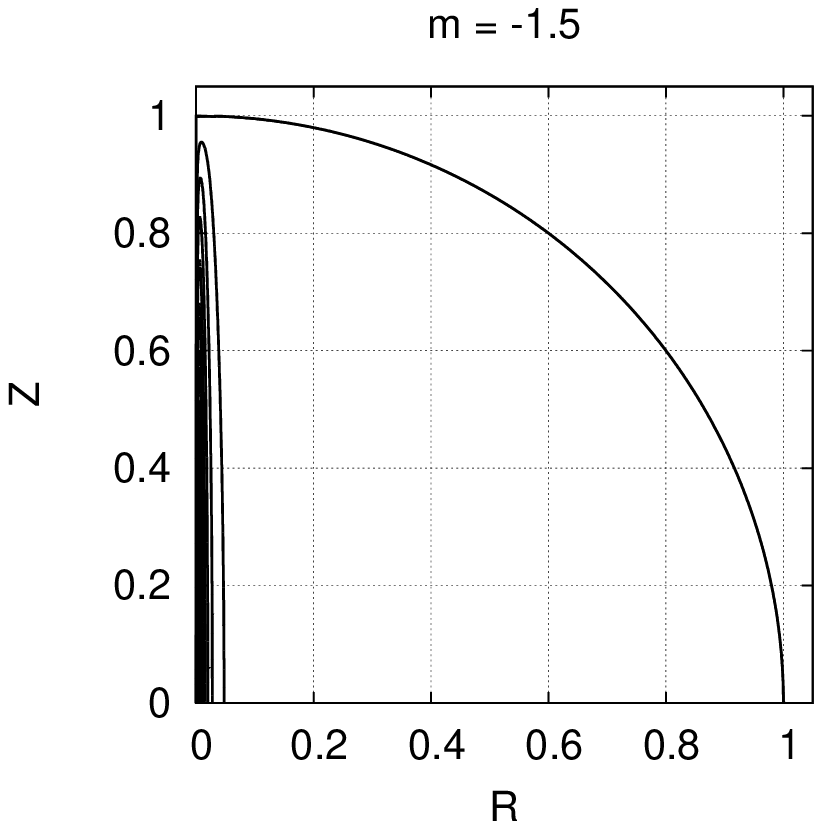}
\includegraphics[scale=0.55]{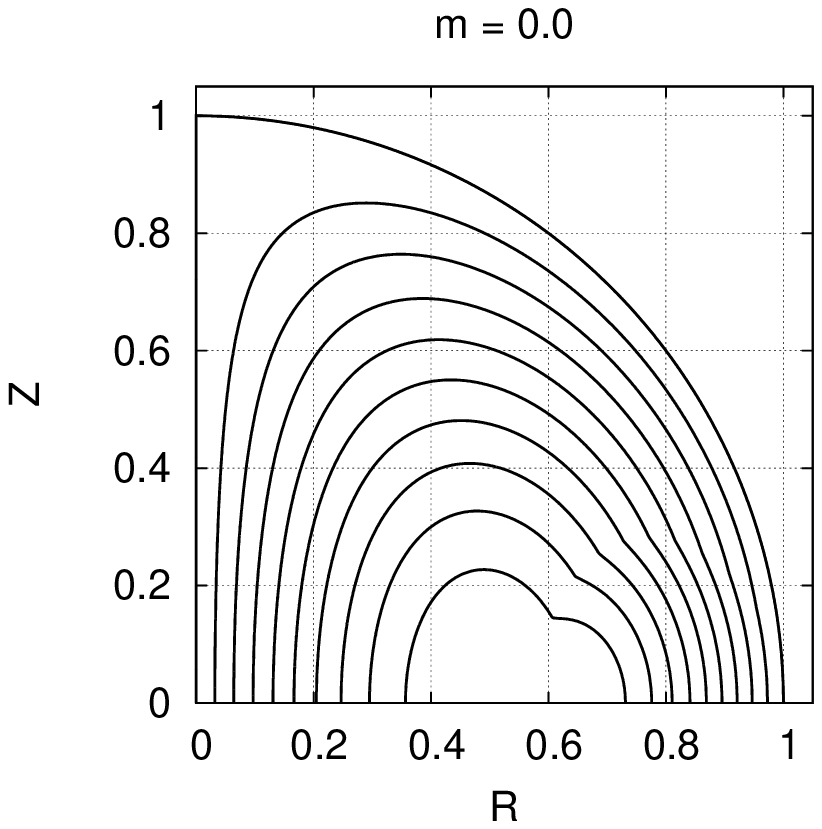}
\includegraphics[scale=0.55]{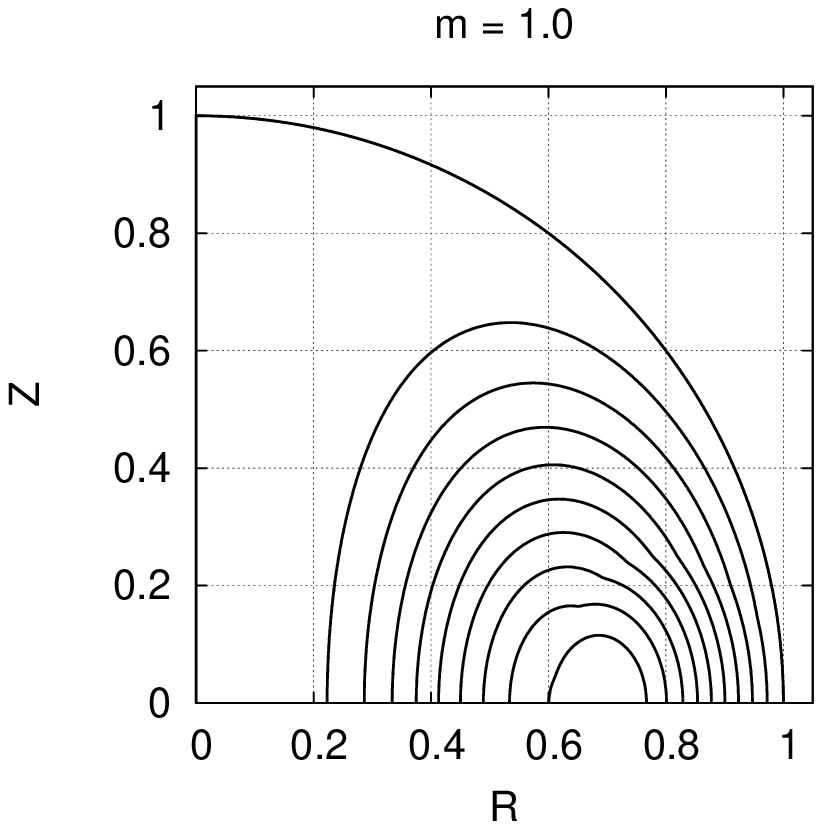}

 \caption{
Isocontours of $j_\varphi$ for different values of $m$. 
These panels show $N=1$, $q=0.99$ polytropic equilibrium models.
The outermost curve denotes the stellar surface. 
The difference between two adjacent contours is 1/10 times 
the maximum of $j_\varphi$.
The current is non-zero in the whole star except at the 
stellar surface and the symmetric axis (z-axis).
}
 \label{fig:j_phi}
\end{minipage}
\end{figure*}

\begin{figure*}
\begin{minipage}{150mm}
\includegraphics[scale=0.43]{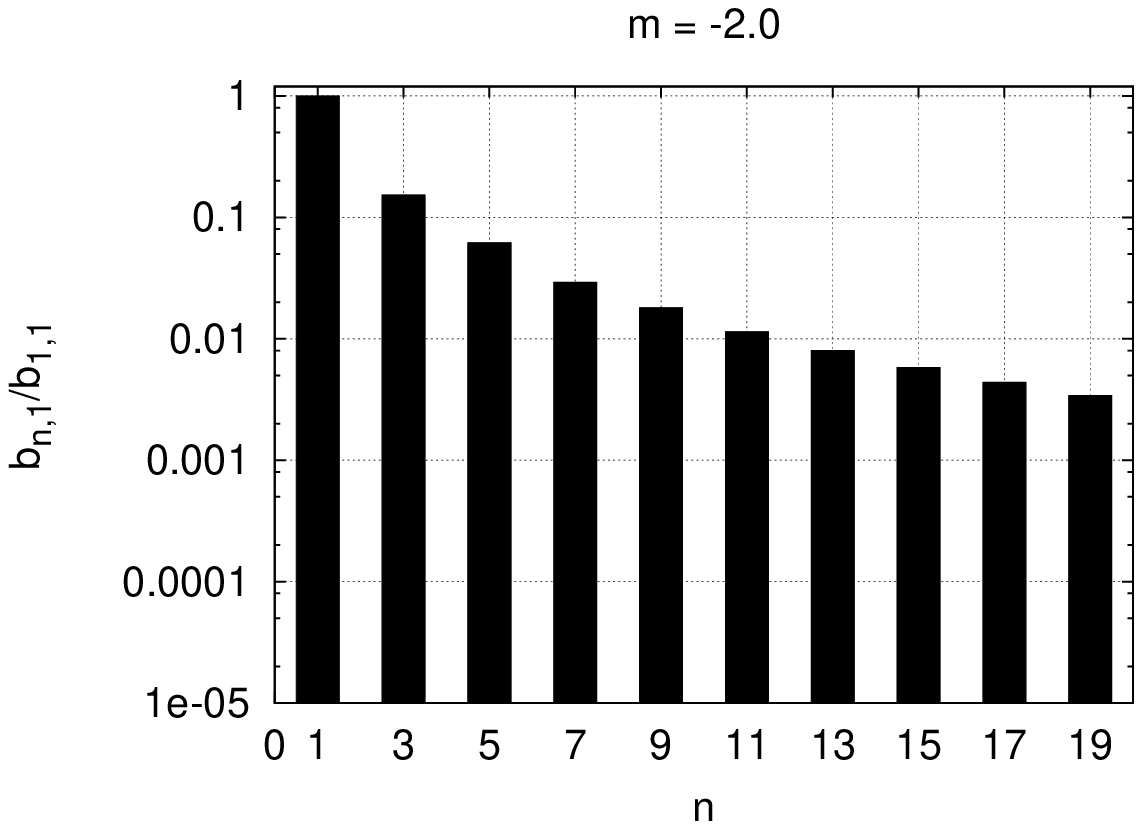}
\includegraphics[scale=0.43]{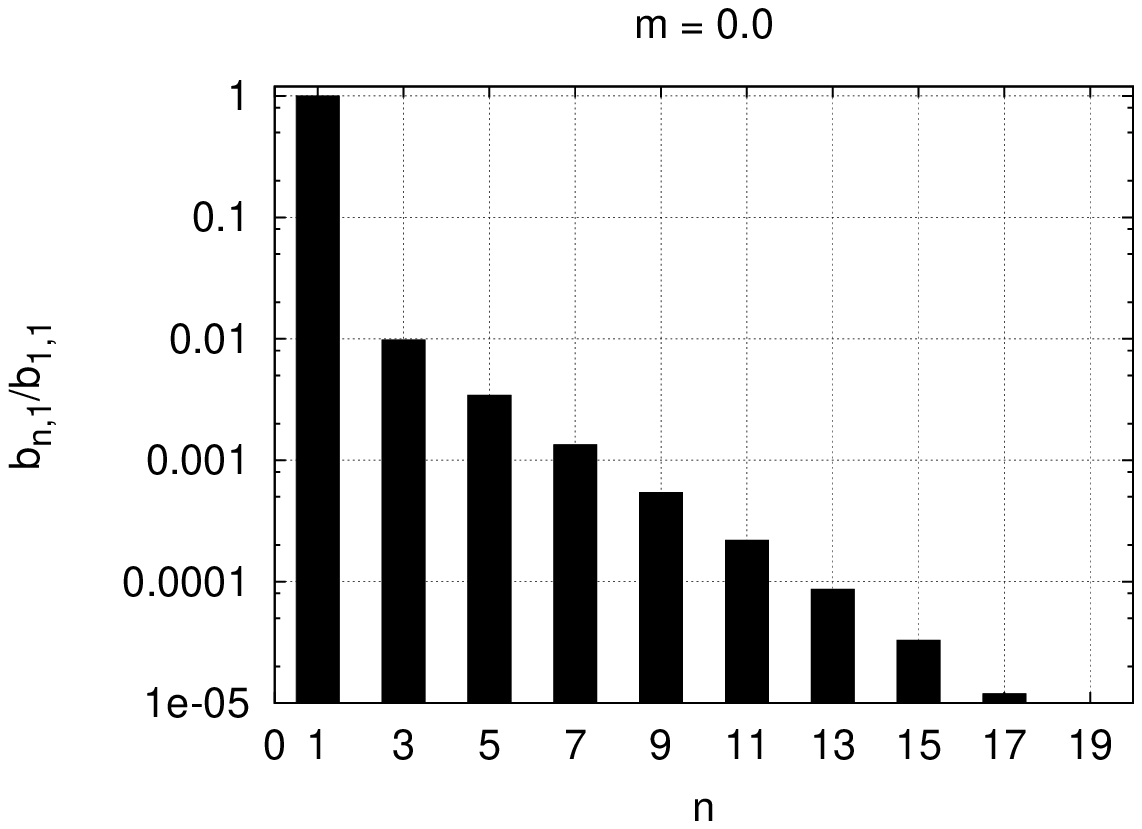}
\includegraphics[scale=0.43]{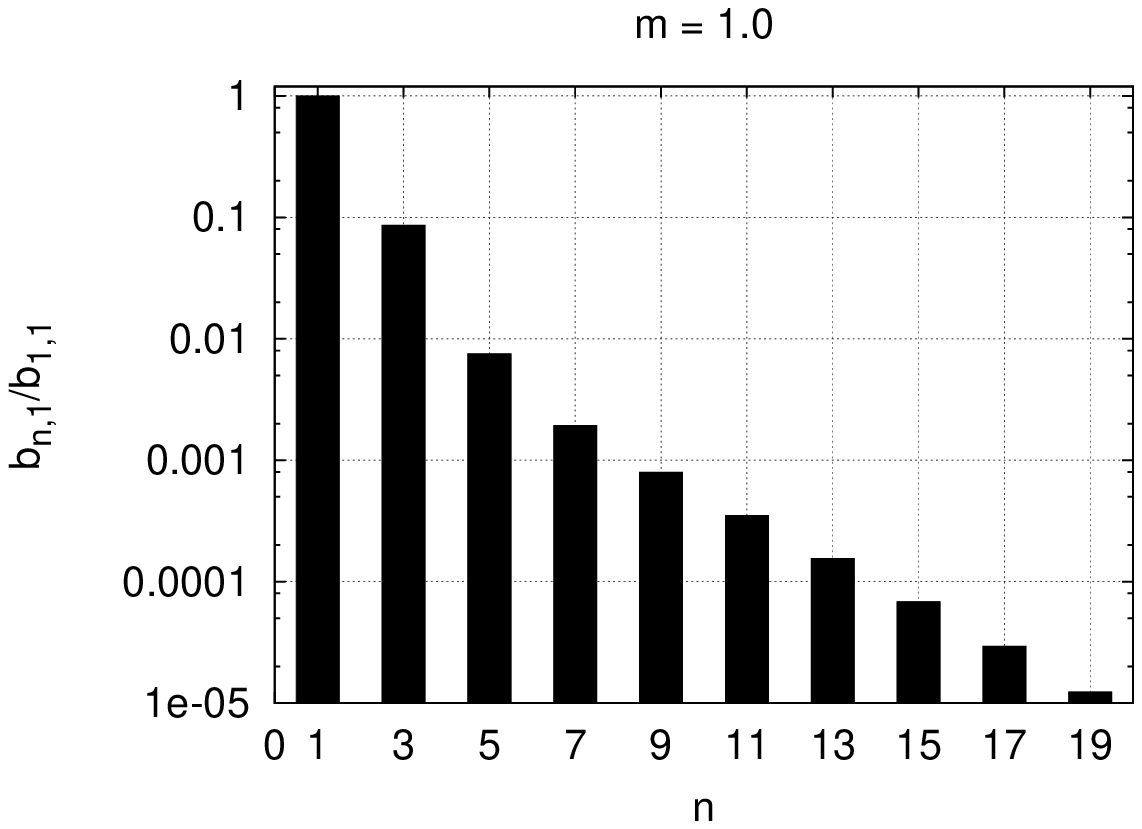}

\caption{The ratio of the magnetic $2^{n}$-pole moment coefficient
to the magnetic dipole moment coefficient,  $|b_{n,1} / b_{1,1}|$,
is plotted against the multipole moment number $n$. Here $b_{n,1}$
is defined in Eq. (\ref{Eq:bn}). 
}
\label{fig:A_phi}
\end{minipage}
\end{figure*}

We show the results for the distributions
of the magnetic fields for different values of $m$.
In particular, in order to examine the effect of  magnetic
fields alone, we consider configurations without rotation. 
The effect of stellar rotation is discussed in 
Sec. \ref{Sec:rotation}. Thus we set $\hat{\Omega}_0 = 0$ and 
compute $N=1$ polytropic equilibrium models with different
values of $ m$ and appropriate values of $q$ so that the 
surface magnetic field becomes roughly $H_{sur} = 10^{15}$  G 
when $\rho_c = 1.0 \times 10^{15} \mathrm{g cm^{-3}}$
and mass $M = 1.4 M_\odot$.
 By setting $N=1$ and an appropriate choice of polytropic 
constant $K$ of $p=K\rho^{2}$, we obtain models with $M=1.4M_\odot$.
It should be noted that these models have
the typical mass and radius for neutron stars. 
We choose $N=1$ as a simple approximation of neutron stars here.
We searched and found the value  of $q$ by calculating 
many equilibrium states.

Physical quantities of these models are shown  in Table 
\ref{tab:m-mu}. It can be seen that values of $\Pi / |W|$ and 
$\alpha$ are almost the same among these models.
Although the strength of the averaged surface magnetic field 
is $H_{sur} = 1.5 \times10^{15}$ G, the values of 
${\cal H} / |W|$ are much smaller than those of  $\Pi / |W|$. 
It implies that the effect of the magnetic fields in these
configurations on their global structures is very small.
On the other hand, values of $H_c / H_{sur}$ and ${\cal H}/ |W|$ vary
rather considerably for different values of $m$.
As the value of $m$ is decreased, values of $H_c / H_{sur}$ and 
${\cal H}/ |W|$ increase.  In Fig. \ref{fig:magnetic_structure} 
the structure and strength of magnetic fields are shown
for three different values of $m$, i.e.  $m = -2.0$ 
(negative $m$ model), $m = 0.0$ ($\hat{\mu}$ = constant model) 
and $m = 1.0$  (positive $m$ model). The left-hand panels show the {\it poloidal} 
magnetic field lines and the regions where the {\it toroidal} magnetic 
field exists. The right-habd panels display the strength of the 
magnetic field $|\Vec{H}|$ normalized by the averaged surface 
magnetic field $H_{sur}$.

As seen from these figures,  there are no discontinuities 
of the magnetic fields at the stellar surfaces.
Due to the choice of the functional form of the 
arbitrary function $\hat{\kappa}(\hat{\Psi})$ and 
the distribution of the magnetic flux function, 
{\it toroidal} magnetic fields appear only in the region
that is bounded by the outermost closed {\it poloidal} 
magnetic field line inside the star 
(thick line). 
Thus the toroidal magnetic fields exist 
inside the torus region.

As the value of $m$ is increased, i.e. from top panels to bottom 
panels, the  ratio of $H_c / H_{sur}$ decreases 
(see left panels in Fig. \ref{fig:magnetic_structure}) 
because the {\it poloidal} magnetic field becomes weaker. 
This is also related to the fact that the interior {\it poloidal} 
magnetic field  lines are much more localized near the axis for 
negative $m$ models. The contours of magnetic field strength 
also display the same tendency.  For the $m = 1.0$ model, the 
contour of $|\Vec{H}| = H_{sur}$ (thick line) shows 
the stellar surface and the shapes of contours are nearly spherical.
By contrast, the contours of the $m=-2.0$ model are highly 
distorted near the axis. The strength of the {\it poloidal} magnetic 
fields for the negative $m$ models could exceed $10^{17}$G near 
the central region. Fig. \ref{fig:beta} shows the profiles of 
the plasma $\beta$ on the $\theta = \pi / 2$ plane, i.e. on the 
equatorial  plane.  Here, the plasma $\beta$ is defined as 
follows:
\begin{eqnarray}
 \beta = 8 \pi p / |\Vec{H}|^2 .
\end{eqnarray}
This quantity denotes the contribution of the gas pressure
effect compared with the magnetic pressure effect.
Fig. \ref{fig:beta} shows profiles of $\beta$ for models with 
$m = -2.0, 0.0, 1.0$.  As seen from Fig.\ref{fig:beta}, 
the profiles of $\beta$ are very similar to each other near the stellar 
surface regions.  
For the region around  $\hat{r} \sim 0.6$, however, the value of 
$\beta$ for  the $m=-2.0$ model is larger than 
those for the $m=0$ and $m=1.0$ models.  Since these models have 
almost the same mass density distributions, 
this difference means a difference of magnetic pressure 
distribution.  In this region  the magnetic field of 
the $m=-2.0$ configuration is weaker and thus the $\beta$ becomes 
larger. However, {\it it should be noted that}
these contours for the model with $m=-2.0$ are 
rather confined to the very narrow region near the central part.  
In other words, the gradient of the magnetic field distribution
for the model with $m = -2$ is much steeper than the 
gradient of the gas pressure distribution compared with
the models with $m=1.0$ and $m = 0.0$. Thus the value 
of $\log \beta$ becomes dramatically small within the $\hat{r} [0:0.1]$
region and the minimum value of $\beta$ can reach about 
$\sim$ 20 in the central part. Therefore, in the central region 
of the model with $m=-2.0$ the influence of magnetic field on 
the local structure of the star is no longer negligible. 

Here we explain the reason why this kind of highly localized 
{\it poloidal} magnetic field configuration can be realized.
We need to note the distribution of the {\it toroidal} current density 
$\hat{j}_\varphi$ in order to analyse our models properly,
because the current density is related to the magnetic field 
closely by the two equations (\ref{Eq:rotH}) and 
(\ref{Eq:current}). In Fig. \ref{fig:j_phi} we show the 
distributions of the {\it toroidal} current density for models
with different values of $m$.  As seen from Fig. \ref{fig:j_phi}, 
the distribution of the {\it toroidal} current density is concentrated 
toward the magnetic axis for the configuration with negative values
of $m$. This is due to the dependence of $\hat{\mu}$ on
the value of $m$. The current density distribution 
spreads over a large region inside
the star as the value of 
$m$ {\it increases} (from left panel to right panel).
In other words, the distribution of the magnetic flux
function becomes more and more concentrated toward the magnetic 
axis as the value of $m$ {\it decreases}.
It implies that the strengths of magnetic fields for models 
with negative values of $m$ become very great near the magnetic 
axis.  Our results show one possibility that a strong {\it poloidal} 
magnetic field can exist deep inside a star. If such a strong 
{\it poloidal} magnetic field is sustained deep inside a star, 
the contours of the magnetic field strength are no longer nearly spherical
as in the bottom right panel of
 Fig. \ref{fig:magnetic_structure}. 
Although this feature might be modified by dropping the assumption of 
the axisymmetry, it would  give us one possibility for 
the presence of a strong {\it poloidal} magnetic field configuration
deep inside a star.

Finally, to characterize the magnetic structure
we show the magnetic multipole moments of magnetized 
stars. In Fig. \ref{fig:A_phi} the values of $|b_{n,1} / b_{1,1}|$ 
(equation \ref{Eq:bn}) are plotted. There appear to be only multipolar magnetic moments 
with odd degree ($n=1, 3, 5$), because we have assumed the 
equatorial symmetry. As seen from these figures, in configurations 
with negative values of $m$ the higher order magnetic multipole 
moments contribute ($|b_{n,1}/b_{1,1}|$) significantly to the total 
magnetic field, while in configurations with positive values of 
$m$ the magnetic dipole moment is the dominant component of the 
total magnetic field.  These figures show that the external magnetic 
field is nearly dipole when we adopt $m=0$ but it is not simple 
dipole when $m > 0$ and $m < 0$. From the left panel, we see 
that the $n=3$ (octupole) component reaches about a few tens of per cent 
of the dipole component when $m=-2.0$.
 
\subsection{Effect of stellar rotation}
\label{Sec:rotation}

We calculate two sequences with rotation for different
values of $m$ 
in order to examine the influence of rotation.  
We choose the value of $\hat{\mu}_0$ by obtaining
a configuration with $\hat{\Omega}_0 = 0$ and $q = 0.99$  
as a non-rotating limit of our equilibrium sequence. 
We choose $q = 0.99$ here for simplicity.
The value of $q = 0.99$ corresponds to an equilibrium 
configuration with $H_{sur} \sim 10^{15}$ 
G when we consider a typical neutron star model with negative $m$.
We have obtained sequences of stationary configurations 
by fixing the parameters $m$ and $\hat{\mu}_0$ and changing 
the value of $q$. By changing the value of $q$ for a
fixed value of $\hat{\mu}_0$, we have equilibrium 
configurations with shapes that are deformed from spheres
by rotational effect in addition to the magnetic force. 
Since we fix the magnetic potential parameter $\hat{\mu}_0$ 
and $m$ along one sequence, the equilibrium 
sequence is the one with approximately constant magnetic effect.
If the values of $m$ and $\hat{\mu}_0$ are changed,
we will be able to solve  another stationary sequence. 
We have calculated two stationary sequences with negative 
$m$ ($m = -1.5$) and with  $m = 0.0$, i.e. $\hat{\mu} = $ constant .


\begin{table*}
\begin{minipage}{150mm}
\hspace{-30pt}
\begin{tabular}{cccccccccccc}
\hline
$q$ & $H_\mathrm{c} / H_\mathrm{sur}$ & ${\cal H}_p /{\cal H}$ &
$|\hat{W}|$ & ${\cal H}/|W|$ & $\Pi/|W|$ & $T/|W|$ &$\alpha$ & $\hat{\Omega}_0^2$
& $\hat{K}$ &  VC\\
\hline
&&&& $m=-1.5$ &&\hspace{-30pt} $\hat{\mu}_0 =$ 5.070E-7 \\
\hline

0.99& 4.75E+1 & 0.9979 & 9.71E-2 & 1.14E-4 &
3.33E-1 & 0.00E+0 & 5.06E-2 & 0.00E+0 & 3.33E-6 & 1.80E-6 \\

0.9 & 4.54E+1 & 0.9972 & 8.00E-2 & 1.26E-4 &
3.19E-1 & 2.14E-2 & 4.51E-2 & 1.17E-2 & 3.61E-6 & 3.41E-5 \\

0.8 & 4.36E+1 & 0.9960 & 6.16E-2 & 1.45E-4 &
3.02E-1 & 4.67E-2 & 3.87E-2 & 2.36E-2 & 3.95E-6 & 1.34E-5 \\

0.7 & 4.19E+1 & 0.9940 & 4.40E-2 & 1.74E-4 &
2.85E-1 & 7.30E-2 & 3.22E-2 & 3.37E-2 & 4.32E-6 & 5.70E-6 \\

\hline
&&&& $m= 0.0$ &&\hspace{-30pt} $\hat{\mu}_0 =$ 5.520E-2 \\
\hline

0.99 & 6.98E+0 & 0.9947  & 9.60E-2 & 2.31E-3 & 
3.33E-1 & 0.00E+0 &  5.03E-2 & 0.00E+0 & 1.20E-4 & 1.85E-6 \\

0.9 &  6.63E+0 & 0.9934 &  7.90E-2 &  2.35E-3 &
 3.18E-1 &  2.14E-2 &  4.48E-2 &  1.16E-2 &  1.15E-4 & 1.24E-6 \\
 
0.8  &  6.29E+0 & 0.9913 &  6.05E-2 &  2.39E-3 &
 3.01E-1 &  4.73E-2 &  3.83E-2 &  2.37E-2 &  1.06E-4 & 1.36E-6 \\
 
0.7 &  5.99E+0 & 0.9878 &  4.32E-2 &  2.39E-3 &
 2.83E-1 &  7.40E-2 &  3.18E-2 &  3.37E-2 &  9.24E-5 & 1.51E-6 \\
 \hline

\end{tabular}
\caption{Physical quantities of two sequences 
with $m = 0.0$ and $m= - 1.5$.}
\label{tab:q-mu}
\end{minipage}
\end{table*}

Physical quantities of stationary configurations 
are tabulated in Table \ref{tab:q-mu}.
As seen from this table, the quantities 
$|\hat{W}|$ and $\alpha$ or the ratio  $\Pi / |W|$ and $T/|W|$ 
depend on the strength of the rotation.
By contrast, magnetic quantities are almost unaffected 
by rotation. The dependence of the ratio $H_{c} / H_{sur}$ on 
rotation is relatively small.  
The equilibrium configurations with highly localized 
magnetic fields that we have obtained in this paper 
 are almost unchanged even by rapid rotation.  
Therefore, we do not consider the effect of rotation any 
longer in this paper.

\subsection{Effect of equations of state}
\label{Sec:EOS}

Thus far, we have discussed our magnetized configurations by 
showing the results for $N = 1$ polytropic models. 
The distribution of the {\it toroidal} current density, however, 
depends on the mass density profile through equation (\ref{Eq:current}). 
Thus we show other polytropic models, i.e. $N=0.5$ 
and $N=1.5$ polytropes, as well as configurations for degenerate
gases, i.e. white dwarf models, in order to examine the influence 
of equations of state on configurations with highly localized 
magnetic fields. 

We set $q=0.99$ for polytropes and $q = 0.999$ for 
degenerate gases. The degenerate model with $q=0.999$ 
corresponds to a configuration with a $H_{sur} \sim 1.0 
\times10^{9}$G magnetized white dwarf with $m = -3.0$, 
the central density is $1.0 \times 10^{8} \mathrm{g  cm^{-3}}$. 
This central density results in a white dwarf of about $1.16 
M_\odot$. Neither models rotates.  We  calculate 11 models 
with fixed values for $q$ by setting $m = $
$-3.0, -2.5, -2.0, -1.5, -1.1, -0.9, -0.5, 0.0, 0.5, 1.0, 
1,3$ and examine the dependence of $H_c / H_{sur}$ on the 
equation of state.


\begin{figure*}
 \begin{minipage}{150mm}
 \begin{center}
  \includegraphics[scale=0.75]{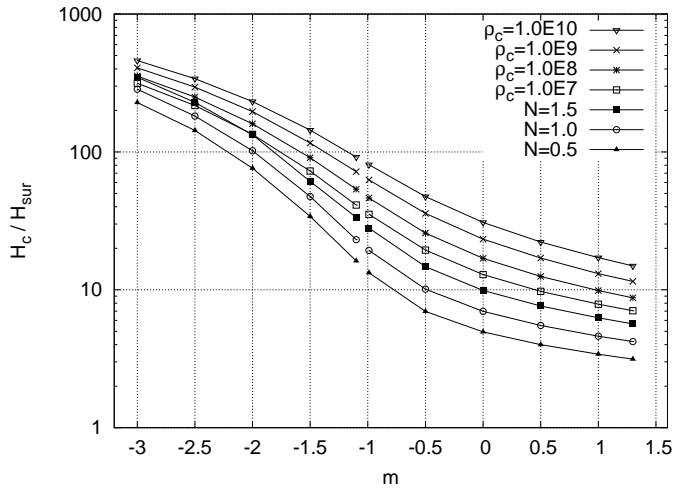}
\caption{The value of $H_\mathrm{c}/H_\mathrm{sur}$ is plotted
against the value of $m$ for polytropes ($q=0.99$) and 
white dwarfs ($q=0.999$).
}
\label{fig:m_ratio}
 \end{center}
 \end{minipage}
\end{figure*}


\begin{figure*}
 \begin{minipage}{150mm}
\includegraphics[scale=0.78]{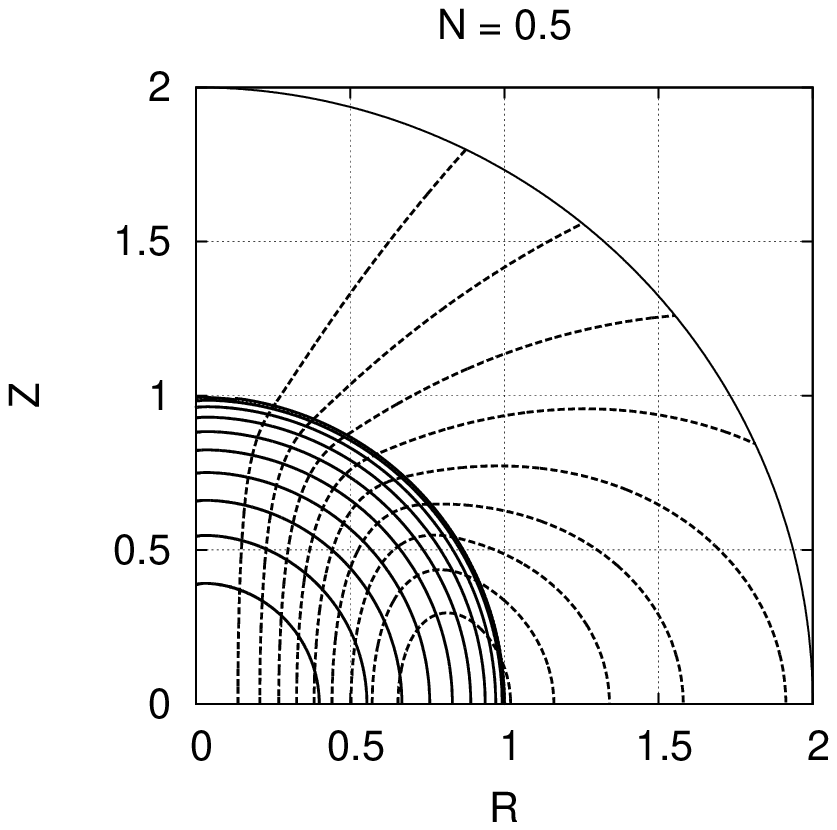}
\includegraphics[scale=0.78]{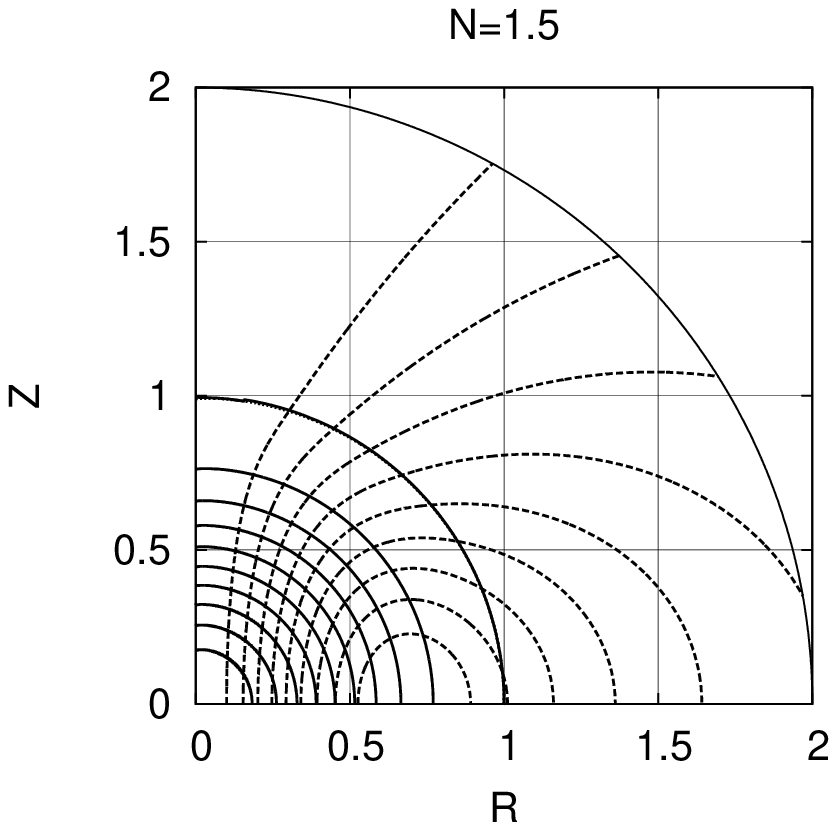}

\includegraphics[scale=0.78]{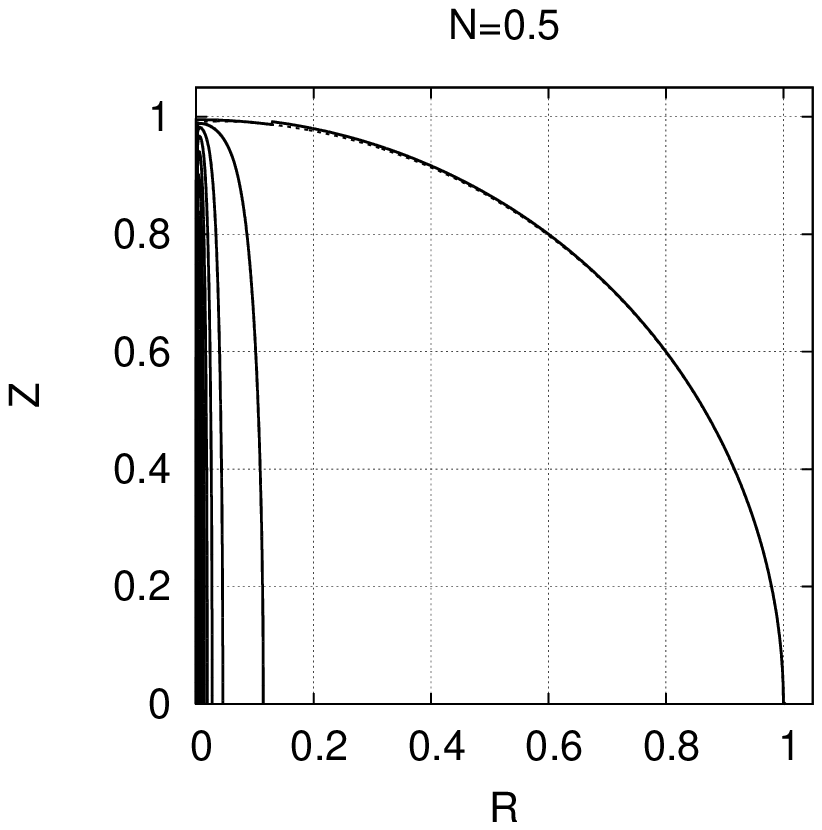}
\includegraphics[scale=0.78]{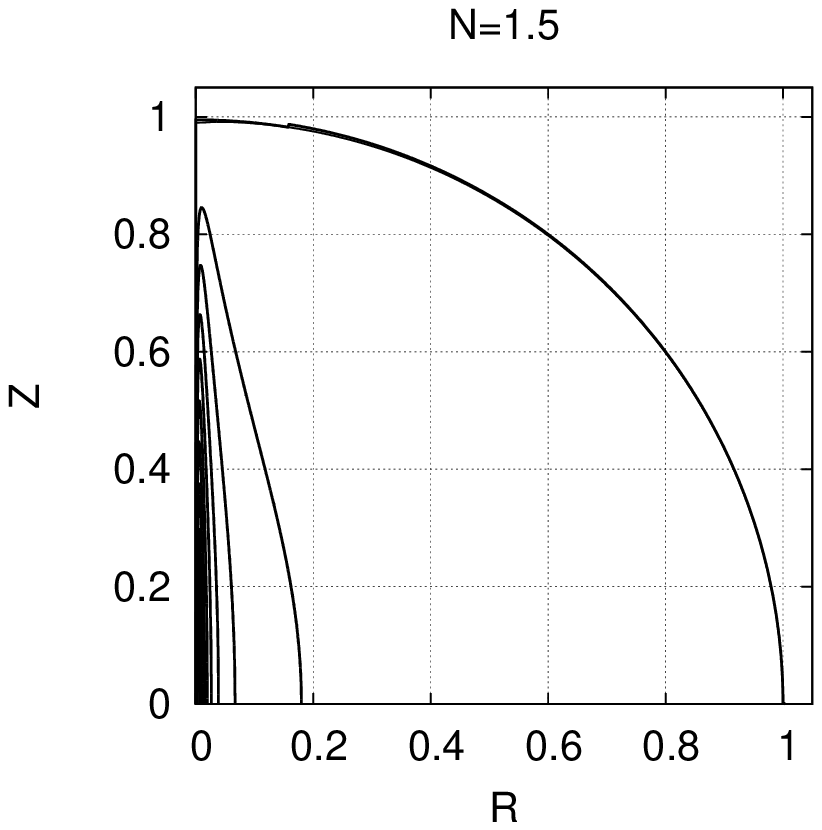}

\includegraphics[scale=0.78]{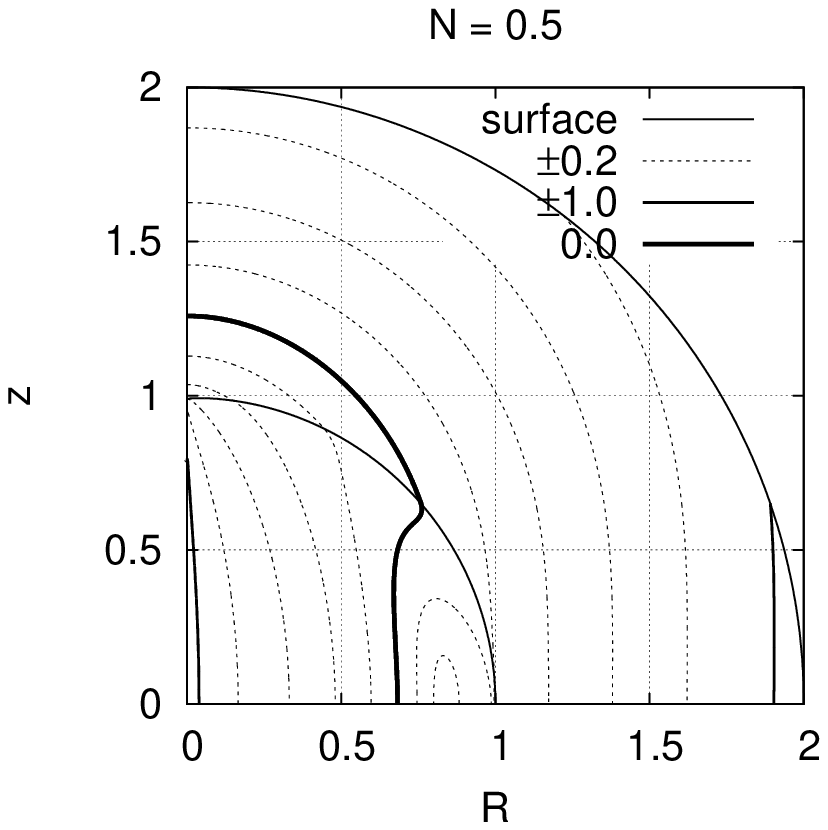}
\includegraphics[scale=0.78]{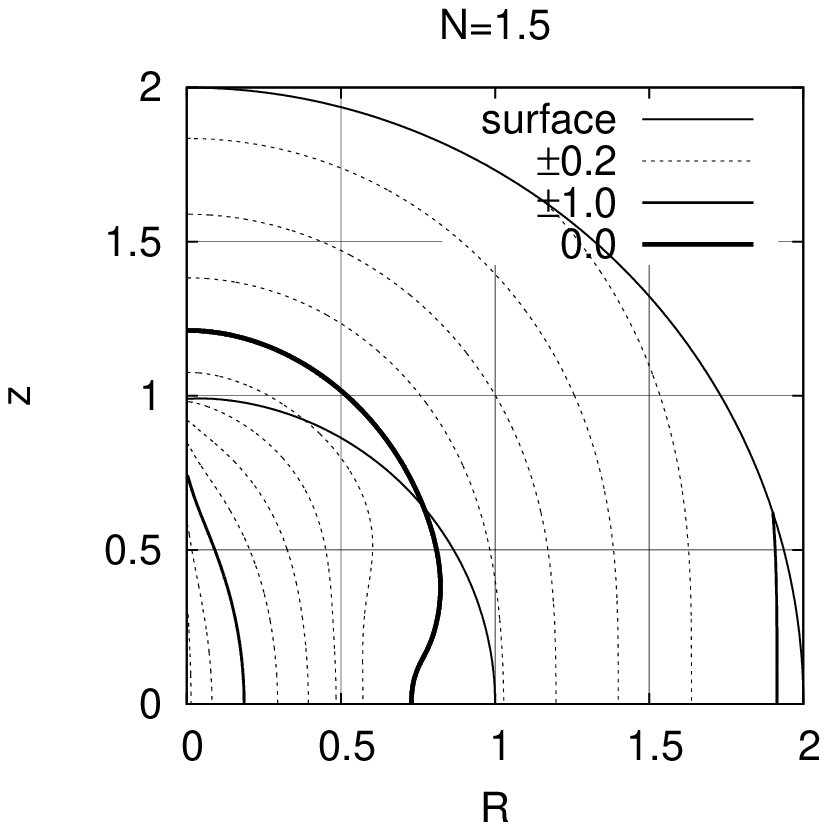}
  
\caption{
Isocontours for $\rho$ and $\Psi$ (top), 
$j_\varphi$ (middle) and $\log [|\Vec{H}|/ |H_{sur}|]$ (bottom).
These panels are for configurations with $m=-0.99$. 
The left panels are contours for the model of a $N=0.5$ polytrope
and the right panels are for a $N=1.5$ polytrope.
 The difference between two 
adjacent contours is 1/10 times the maximum of the corresponding
quantities.
}
\label{fig:rho1}
 \end{minipage}
\end{figure*}

\begin{figure*}
 \begin{minipage}{150mm}
  
\includegraphics[scale=0.78]{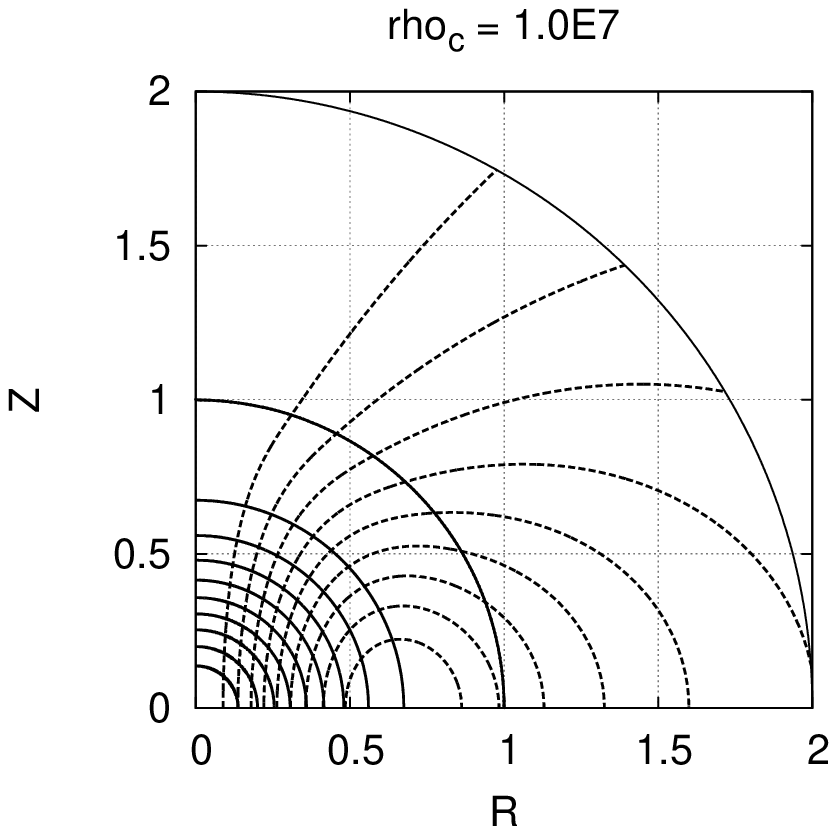}
\includegraphics[scale=0.78]{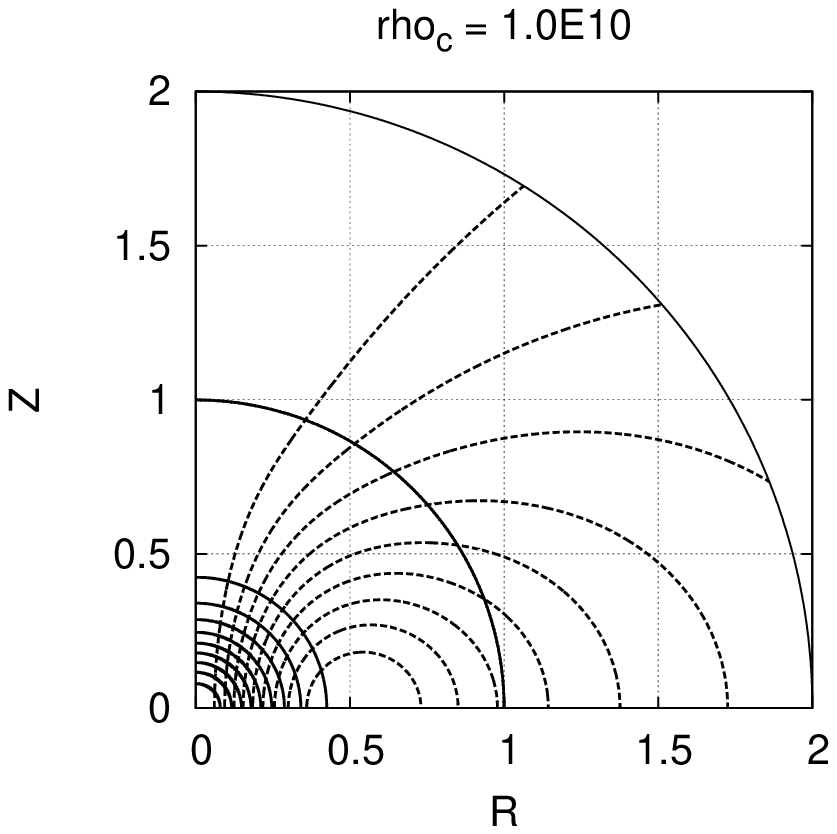}

\includegraphics[scale=0.78]{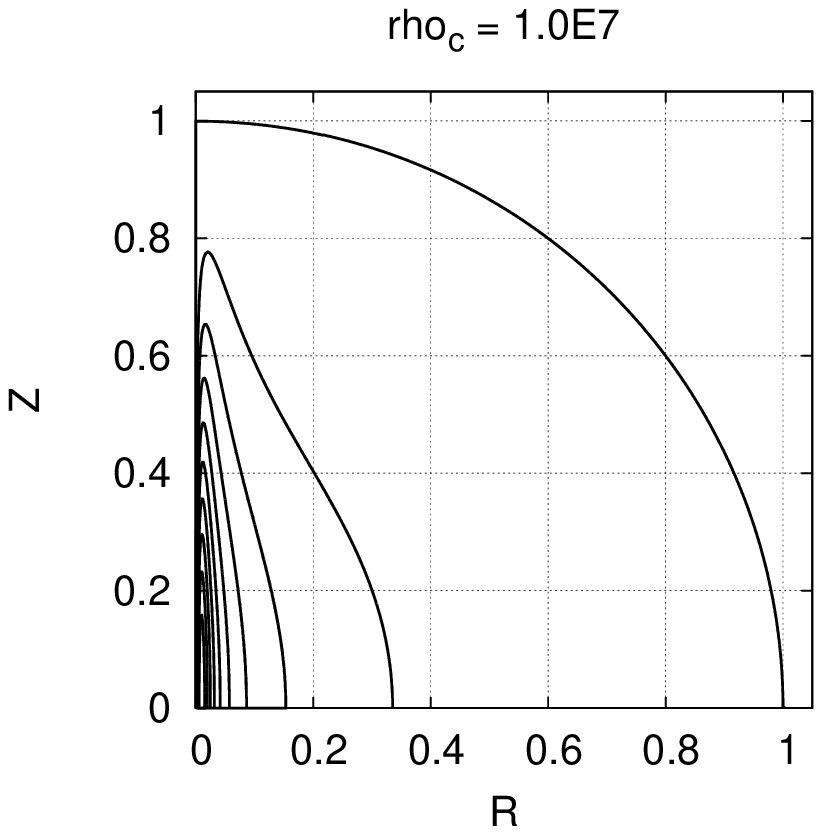}
\includegraphics[scale=0.78]{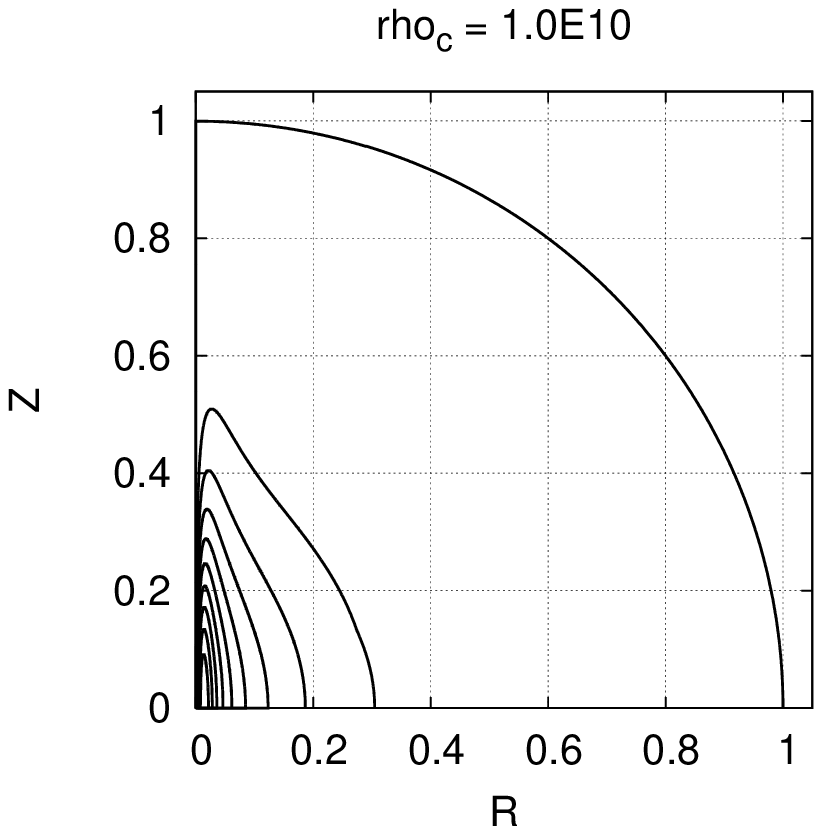}

\includegraphics[scale=0.78]{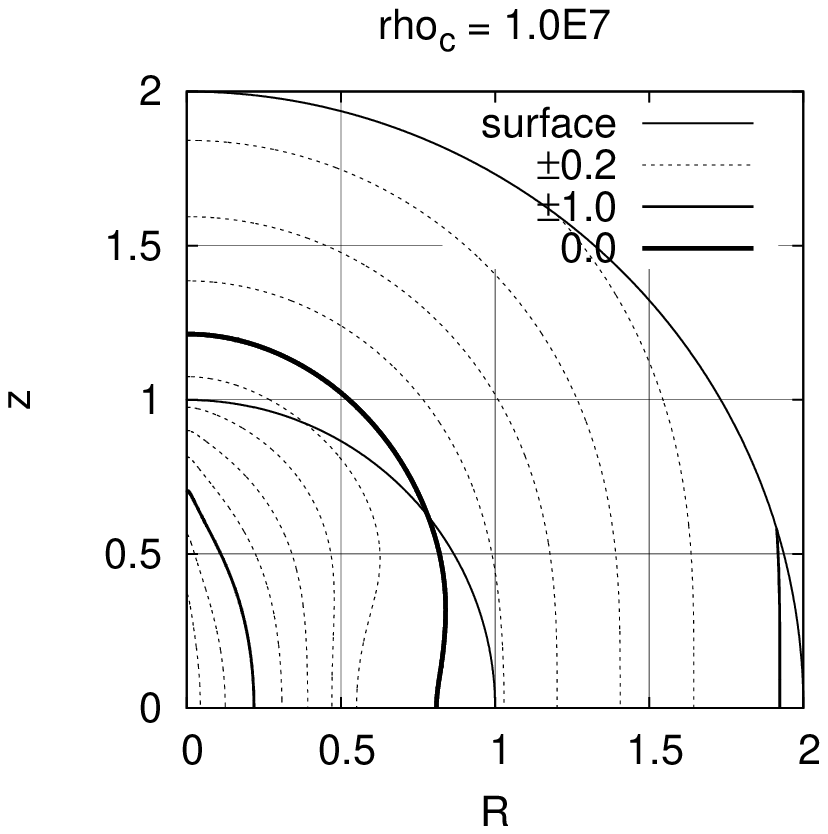}
\includegraphics[scale=0.78]{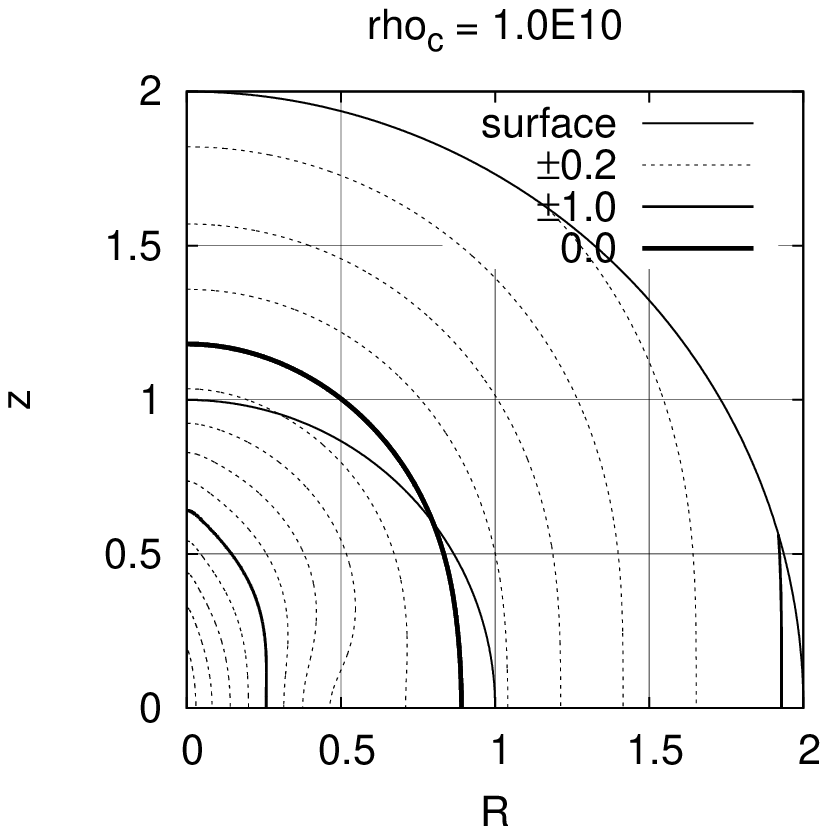}

\caption{
Same as Fig.\ref{fig:rho1} except for the 
equations of state. 
The left panels are for a white dwarf with
$1.0\times 10^{7} \mathrm{g cm^{-3}}$
and the right panels are for a white dwarf with $1.0\times 10^{10} \mathrm{g  cm^{-3}}$ .
}
\label{fig:rho2}
 \end{minipage}
\end{figure*}

Fig. \ref{fig:m_ratio} displays the ratio $H_c / H_{sur}$ 
against the value of $m$ for
different equations of state. The dependency of this 
ratio on the value of $m$ is qualitatively similar for these equations 
of state. Whichever equation of state  we choose, we obtain 
configurations with highly localized magnetic fields, for which
$H_c / H_{sur}$ can exceed 100. The same is true for 
white dwarfs with highly localized magnetic fields. 
However, $H_c / H_{sur} $ tends to become smaller for
stiffer equations of state,  as seen from Fig. \ref{fig:m_ratio}. 

Fig. \ref{fig:rho1} and Fig. \ref{fig:rho2}
display  the distribution of mass density,
current density and the contour of $\log_{10}|\Vec{H}| / H_{sur}$ 
of $m = -0.99$ configurations.
Fig. \ref{fig:rho1} shows results for polytropes
$N=0.5$ and $N=1.5$ (stiffest and softest equations 
of state among the polytropic models considered here)
and Fig. \ref{fig:rho2} shows results for  white dwarfs
with $\rho_c = 1.0 \times 10^{7} \mathrm{g cm^{-3}}$ 
and $\rho_c = 1.0 \times 10^{10} \mathrm{g cm^{-3}}$ 
(stiffest and softest among the white dwarf models considered here).
As seen from top panels in each figure, the 
mass density distributions of the softer equation of state
($N=1.5$ and $\rho_c = 1.0 \times 10^{10} \rm{g cm^{-3}}$)
 are  more centrally concentrated than those of the stiffer 
equation of state ($N=0.5$ and $\rho_c = 1.0 \times 10^{7} \rm{g cm^{-3}}$). 
The current density distributions are also more centrally
concentrated compared with the mass density distribution  
(middle panels). As a result, the {\it poloidal} magnetic fields 
become more highly localized for the softer equation of 
state (bottom panels). The mass of the white dwarf becomes 
higher for the  higher central density. This implies that 
higher mass white dwarfs can have 
stronger  interior magnetic fields deep inside
if the magnetic field structure is fixed as in the present study.

\section{Discussion and conclusions}

In this paper we have constructed axisymmetric and stationary 
magnetized barotropes that have extremely strong 
{\it poloidal} magnetic fields around the central region 
near the magnetic axis. The strength of the magnetic field 
in that region could be two orders of magnitude
larger than that of the surface magnetic field. In the 
context of the neutron star physics, this would imply that 
there might be magnetars whose interior magnetic fields 
amounting to $10^{17}$ G if we assume the surface field to 
be order of $10^{15}$G and  that there might be magnetized white dwarfs with
interior magnetic fields that reach $10^{12}$ G when the mass is 
nearly the Chandrasekhar limit and the surface field is of the 
order of $10^9$ G.

Moreover,  it should be noted that highly localized 
magnetized stars could have higher order magnetic multipole 
moments in addition to the dipole moment.
Although in most astrophysical situations magnetic 
dipole fields have been assumed, we may need to consider 
configurations with contributions from higher multipole
magnetic moments for some situations. In those cases, 
configurations with negative values of $m$ might 
be used to analyze  such systems. 

\subsection{Higher order magnetic multipole moments with even $n$ }

It should be noted that in the analysis of this paper only
higher magnetic multipole moments with odd $n = 2 \ell + 1$
where $\ell$ is an integer, i.e. $2^{2 \ell + 1}$
moments, appear and that there are no higher magnetic 
multipole moments with even $n = 2 \ell$. This is due to the
choice of the current density. Our choice of the arbitrary
function $\mu(\Psi)$ and the assumption of the symmetry of 
$\Psi$ about the equator necessarily result in  magnetic 
field distributions that are symmetric about the equator. 
It implies that the magnetic field should penetrate the equator 
and that $2^{2 \ell}$ type distributions that are confined 
the upper or lower half of the space of the equator
are excluded. In order to obtain closed magnetic field distributions
in the half plane above or below the equator, the current density must be chosen so as to flow
in opposite directions above and below the equatorial plane.
It also implies that we need to set the current density on the
equator in the $\varphi$-direction to vanish. 

Concerning $2^{2 \ell}$ multipole magnetic moments 
distributions, \cite{Ciolfi_et_al_2009} have obtained
such configurations. Their solutions correspond to the
choice of the current density distributions that
are antisymmetric about the equator.

\subsection{Forms of arbitrary functions}

One might think it curious that 
functions appear in the formulation and that there is no physical
principle specifying  how to choose those arbitrary functional 
forms. The same situation appears for the problem
of calculating equilibrium structures or stationary
structures of rotating and axisymmetric 
{\it barotropes}. For that problem, the
three component equations  of the equations of motion 
do not remain independent but come to depend on each other.
This implies that one could not solve for all the three components of
the flow velocity completely. Assumptions of the 
{\it stationarity} and {\it barotropy} reduce the
problem to a degenerate problem concerning the components of the 
flow velocity. Although there are {\it three} component 
equations for the {\it three} components of the flow velocity, 
those three component equations are no  more independent.
They become dependent each other due to the {\it barotropic
nature} of the assumption for the gas. Therefore, one
needs to {\it specify} the {\it rotation law or  corresponding 
relation} in order to find stationary or equlibrium configurations 
for axisymmetric barotropes. The form of the rotation law is {\it arbitrary}.

The only requirement for the functional form regarding  
the rotation law comes from the nature of the stability of the 
system. However, one needs to know the stability of the
system beforehand. If one does not have any 
information about the system to be solved, one has
no principle by which to choose the form of the rotation
law.

The situation is the same for the 
{\it stationary} problem for axisymmetric 
magnetized {\it barotropes}.  For the stationary states of
axisymmetric magnetized barotropes, the situation is 
more complicated than that for rotating barotropes, 
because not only the flow velocity but also the 
magnetic field appears in the problem. That
also leads to the appearance of a greater number of arbitrary 
functions in the problem. Thus it is very hard to
specify the forms of arbitrary functions 
{\it physically meaningfully}. In such situations
the only thing one can might be to explore
many kinds of arbitrary functions to find out
the general consequences of the resulting magnetic fields. 

Of course,
if one could obtain a lot of information of the
magnetic characteristics about the equilibrium
states at hand, one could constrain the arbitrary 
functions more appropriately and more physically 
meaningfully. One possibility is to rely on
the stability nature of the equilibrium,  as in the rotating
barotropic stars. Since there is no useful stability criterion
for the field configuration with both {\it poloidal} and {\it toroidal} fields
and linear stability analysis of the equilibrium is
beyond our scope, we leave this issue of constraining
the functional form for a future study.

\subsection{Application to magnetars}

The typical strength of the surface magnetic field of anomalous X-ray pulsar (AXP) and 
soft gamma-ray repeater (SGR) is considered to be $10^{14}-10^{15}$G by assuming the
magnetic dipole spin down (see e.g. \citealt{Kouveliotou_et_al_1998};
\citealt{Kouveliotou_et_al_1999}; \citealt{Murakami_et_al_1999};
\citealt{Esposito_et_al_2009_mn};
\citealt{Enoto_et_al_2009_apj}; \citealt{Enoto_et_al_2010_pasj}).
According to recent observational evidences, 
some types of AXP and SGR are regarded as similar kinds of isolated 
neutron star and are categorized as magnetars, 
although they were first considered to belong to two
different types of neutron star. (see e.g. 
\citealt{Duncan_Thompson_1992_ApJ};
\citealt{Duncan_Thompson_1996_AIPConf}; 
\citealt{Woods_Thompson_2006};
\citealt{Mereghetti_2008_aar}).

For neutron stars with a strong magnetic field,  such as 
magnetars, the strength of the maximum {\it toroidal}
magnetic field inside has been estimated to be
$10^{17}$ G (see e.g. \citealt{Thompson_Duncan_1995_mn};
\citealt{Kluzniak_Ruderman_1998_apjl}; \citealt{Spruit_1999};
\citealt{Spruit_2009_IAUsymp}).  Many authors have considered
that only  {\it toroidal} magnetic fields could become extremely
strong and be hidden below the surfaces of the stars.
Concerning  {\it poloidal} magnetic fields,  a very strong field
is not considered because it would be observed as a strong
surface field since it is dipole-dominated.
However, as shown in this paper, extremely strong {\it poloidal}
magnetic fields can exist in the very central region at
$r_c \sim 0.01 r_e$,  as seen from Tables \ref{tab:m-mu}
and \ref{tab:q-mu} and Fig. \ref{fig:m_ratio} and the definition
of $H_c$, Equation(\ref{Eq:H_c}). If we apply our equilibrium models
with negative values of $m$ to magnetars with mass 
$1.4M_\odot$, central density $\rho_{\max} = 1.0\times10^{15} 
\mathrm{g cm^{-3}}$ and average strength of the surface magnetic 
fields $10^{15}$ G, the strengths of the {\it poloidal} magnetic 
fields could be  $10^{16} - 10^{17}$G.
We also consider weak magnetized magnetars with
average strength of the surface magnetic fields $10^{13}$ G 
(\citealt{Rea_et_al_2010}). If we apply our equilibrium models, 
the strengths of the {\it poloidal} magnetic fields could be  
$10^{14} - 10^{15}$G. Since these strong {\it poloidal} magnetic fields 
located nearly along the magnetic axis
in the central core region, the magnetic structures in the core
region are highly anisotropic.
If extremely strong magnetic {\it poloidal} fields
are hidden within the core region, there could be magnetic fields
with higher order multipole moments.

If the neutron star shape is 
deformed by a strong magnetic field and the magnetic axis is
not aligned the rotational axis, gravitational waves 
will be emitted (\citealt{Cutler_2002};
\citealt{Haskell_Samuelsson_Glampedakis_2008};
\citealt{Mastrano_et_al_2011}).
Gravitational wave emission tends to become stronger as the ellipticity 
of the meridional plane
of the star becomes larger. 
For our models, decreasing $m$ increases the value of $1-q$
in the $H_{sur}$ constant sequence
(see the value of $1-q$ in Table \ref{tab:m-mu}).
Thus those models with highly localized magnetic field here 
may be efficient emitters of gravitational wave.

\subsection{Some features of highly magnetized white dwarfs}

It is widely believed that the effect of the stellar
magnetic fields play a significant role in astrophysics.
For example, isolated magnetized white dwarfs tend to 
have a higher mass than non-magnetic
white dwarfs (\citealt{Wickramasinghe_Ferrario_2000}).
According to observations, the surface magnetic field strength 
of white dwarfs varies from very little to $10^9$ G 
(\citealt{Wickramasinghe_Ferrario_2000}). 
Therefore, there are some strongly magnetized white dwarfs 
whose surface magnetic field about $10^8$-$10^9$G.
For example, \cite{Jordan_et_al_1998} estimated the field range
$3.0 \times 10^8$-$7.0 \times 10^8$ G in GD 299.
EUVE J0317-855 is a massive high-field magnetic white dwarf
with rapid rotation. Its magnetic field was calculated  
by an offset dipole model with $4.5 \times 10^8$G and 
period of 725 s. PG 1031+234 is a high-field magnetized white dwarf.
\cite{Schmidt_et_al_1986} and \cite{Latter_Schmidt_Green_1987}
estimated its rotation period 3.4 h and
its magnetic field as $5.0\times 10^{8}$ - $1.0 \times 10^9$ G.
The observed spectral variations cannot be fitted well 
by a simple dipole magnetic or offset dipole model, so
they have proposed a two-component model 
composed of a nearly centered dipole and 
a strongly off-centered dipole. 
In other words, the magnetic field 
structures of several strongly magnetized white dwarfs
could not be explained by applying simple dipole structures.

We have obtained strongly magnetized white dwarfs with 
higher order magnetic multipole moments in this paper. If we apply our
configurations with negative $m$, 
some strongly magnetized star such as PG 1031+234 
may have strong interior magnetic fields.
According to our numerical results, 
$H_{c}$ could reach  as high as $10^{12}$ G 
when $H_{sur} \sim 3.0 \times 10^{9}$ G for a  highly localized 
($m = -3.0$) and high mass ($\rho_c = 1.0 \times 10^9$, 
$M \sim 1.34 M_\odot$) model (see Fig. \ref{fig:m_ratio}).  
Since the central magnetic field strength $H_c$ depends on 
the equation of state as we have shown  in Sec.\ref{Sec:EOS}, 
it becomes higher as the central density increases. 
Thus high mass white dwarfs could  have strong {\it poloidal} magnetic 
fields according to our models with negative $m$. 
As we have displayed in Sec. \ref{Sec:current},
$N=1.5$ polytropes  with negative values of $m$ have
rather large higher order magnetic multipole moments.
The same is the case for magnetized white dwarf models,
i.e. they have rather large higher order magnetic multipole 
moments. Therefore,  the magnetic fields outside of such stars 
are far from simple dipole fields if the magnetized white 
dwarfs have highly localized strong {\it poloidal} 
magnetic fields deep inside the stars.

\subsection{Comments on stability of magnetized barotropes}

Once equilibrium configurations are obtained, it would be 
desirable to investigate their stability. However, a 
satisfactory formulation for the linear stability analysis 
for general magnetic configurations has not been fully 
developed, although there is a stability criterion only for
purely {\it toroidal} magnetic configurations 
(\citealt{Tayler_1973_mnras}).
For purely {\it poloidal} or mixed {\it poloidal-toroidal} 
magnetic configurations,  magnetic configurations with 
rotation or other general situations,
no authors have ever succeeded in obtaining a clear 
stability criterion
(see e.g. \citealt{Markey_Tayler_1973_mnras}; 
\citealt{Wright_1973_mnras};
\citealt{Markey_Tayler_1974_mn}; \citealt{Tayler_1980_mn}
; \citealt{BonannoUrpin_2008_AA}).
Therefore the stability of the configurations obtained in 
this paper contain both  {\it poloidal} and 
{\it toroidal} magnetic fields has not been investigated.

By contrast, the stability of magnetized stars may be
investigated through that time-dependent evolutionary computations of 
the magnetic configurations. Thanks to powerful computers, some 
authors have recently employed magnetohydrodynamical codes 
to follow the time evolutions of magnetized configurations and 
find out whether these configurations would
settle down to certain 'stable equilibrium states'.
Such investigations concerning the magnetic configurations have been
carried out by Braithwaite and his coworkers as mentioned
in Introduction (see e.g. \citealt{Braithwaite_Spruit_2004};
\citealt{Braithwaite_Nordlund_2006}; \citealt{Braithwaite_Spruit_2006};
\citealt{Braithwaite_2006}; \citealt{Braithwaite_2007};
\citealt{Braithwaite_2009}; \citealt{Duez_Braithwaite_Mathis_2010}). 
According to their results, purely {\it toroidal}
configurations and purely {\it poloidal} configurations are shown 
to be all unstable, as previously shown or expected 
(e.g. \citealt{Tayler_1973_mnras};
\citealt{Markey_Tayler_1973_mnras}; 
\citealt{Flowers_Ruderman_1977_apj}. 
However, see \citealt{Geppert_Rheinhardt_2006} for some
results about stability).
Concerning the mixed {\it poloidal-toroidal}
magnetic configurations, 
recent numerical studies (\citealt{Braithwaite_2009}; 
  \citealt{Duez_Braithwaite_Mathis_2010}) have shown  
that they are stable as long as the following condition is 
satisfied:
\begin{eqnarray}
    \alpha_0 \frac{{\cal H}}{|W|} \le \frac{{\cal H}_p}{{\cal H}} \le 0.8 \ ,
\label{criterion_by_braithwaite}
\end{eqnarray}
where $\alpha_0$ is a numerical factor of $10 - 10^3$ 
depending on the stellar structures.
By performing 3D MHD simulations, it has been
shown that non-axisymmetric perturbations to equilibrium stars
grow when this condition is not satisfied. Stars with mixed 
magnetic fields whose dominant component is {\it poloidal} field seem
to evolve toward non-axisymmetric 
configurations until the amplitude of the perturbations
reach nonlinear regime and saturate.
As can be seen from tables in this paper, we havefound 
no models that satisfy that criterion 
(equation \ref{criterion_by_braithwaite}) for our particular choice of 
functional forms presented above (see Sec.\ref{sec:bc}),
because the energy stored in the {\it toroidal} magnetic field 
is at most a few per cent for all of our models. 
In order to obtain configurations that satisfy the criterion, 
we need to choose different functional
forms from those used in this paper. 
We should be careful to apply the criterion, however, to general 
configurations of magnetic fields. The class of solutions with 
both {\it toroidal} and {\it poloidal} magnetic fields obtained here may be 
rather different from the ones studied by
Braithwaite and his collaborators, even if they share the obvious
characteristics of twisted-torus structures of  magnetic fields.
As is seen in completely different stability natures of seemingly 
similar configurations in \cite{Geppert_Rheinhardt_2006} and
\cite{Braithwaite_2007}, it is quite uncertain at this moment 
that failure to satisfy the criterion 
(equation \ref{criterion_by_braithwaite}) for our models here 
means unstable nature of them. It would be interesting to study 
the stability nature of our configurations thorough either linear 
perturbation analysis or direct MHD simulations.

\subsection{Conclusions}

In this paper, we have presented an extended formulation for obtaining
axisymmetric and stationary barotropic configurations with both the
{\it poloidal} and {\it toroidal} magnetic fields. 
We have shown the possibility that magnetized stars have
strong {\it poloidal} magnetic fields inside the star.
Our findings and conjectures can be
summarized as follows.
\begin{enumerate}
  \item By choosing the functional form for one of the  arbitrary
   functions that appear in the basic formulation for the configurations
   under the assumptions mentioned before, we have obtained magnetized
   configurations in which extremely strong {\it poloidal} fields are confined
   within the central part of the near axis region. When we apply our models
   to magnetars, the interior magnetic strength would be $10^{17}$ G while the
   surface magnetic strength is $10^{14}$ - $10^{15}$ G. On the other hand,
   if we apply our models to magnetized white dwarfs with mass $\sim 1.34 M_\odot$,
   the surface field strength would be $10^{9}$ G and $H_c$ reaches $10^{12}$ G.

 \item If stars have extremely strong {\it poloidal} magnetic fields 
       deep inside,
       the contours of magnetic field strengths are not 
       spherical but rather column-like shapes as shown in the figures. 

 \item If stars have extremely strong magnetic fields deep 
       inside, contributions from higher order magnetic multipole 
       moments to the outer fields around the stars cannot 
       be neglected. This implies that if stars have highly 
       localized and extremely strong magnetic fields
       deep inside, then observations of magnetic fields around
       the stars could not be explained by the simple dipole models
       that have been used in most situations.
\end{enumerate}

\subsection*{ACKNOWLEDGMENTS}

We would like to thank Dr.~R.~Takahashi for his discussion 
while we were extending the formulation of the present paper.
KF would like to thank Dr.~K.~Taniguchi for discussion and comments 
on this paper.  We would also like to thank the anonymous reviewer for 
useful comments and suggestions that help us to improve this 
paper. This research was partially supported by the
Grant-in-Aid for Scientific Research (C) of Japan Society
for the Promotion of Science (20540225) and 
by Grand-in-Aid for JSPS Fellows.

\bibliographystyle{mn}


\appendix


\section{Numerical method}

\subsection{Dimensionless quantities}
\label{App:dimensionless}

In this paper, physical quantities are used in their
dimensionless forms as follows:
\begin{eqnarray}
   \hat{\phi}_g \equiv \frac{\phi_g}{4 \pi G r_e^2 \rho_{\mathrm{max}}} \ ,
\end{eqnarray}
\begin{eqnarray}
   \hat{\Omega} \equiv \frac{\Omega}{\sqrt{{4 \pi G \rho_{\mathrm{max}}}}} \ ,
\end{eqnarray}
\begin{eqnarray}
 \hat{\kappa} \equiv \frac{\kappa}{\sqrt{4\pi G} r_e^2 \rho_{\max}} \ ,
\end{eqnarray}
\begin{eqnarray}
 \hat{\mu} \equiv \frac{\mu}{\sqrt{4 \pi G}/r_e} \ ,
\end{eqnarray}
\begin{eqnarray}
   \hat{H}_{suffix} \equiv \frac{H_{suffix}}{\sqrt{4 \pi G }
   r_e \rho_{\max} } \ ,
\end{eqnarray}
\begin{eqnarray}
   \hat{A}_\varphi \equiv \frac{A_\varphi}{\sqrt{4 \pi G }r_e^2 \rho_{\max} } \ ,
\end{eqnarray}
\begin{eqnarray}
 \hat{\Psi} \equiv \frac{\Psi}{\sqrt{4 \pi G }r_e^3 \rho_{\max} } \ ,
\end{eqnarray}
\begin{eqnarray}
   \hat{K} \equiv \frac{K}{4 \pi G r_e^6 \rho_{\max}^2 } \ ,
\end{eqnarray}
\begin{eqnarray}
   \hat{j}_\varphi \equiv \frac{j_\varphi}{\sqrt{4\pi G}\rho_{\max} c} \ .
\end{eqnarray}
\begin{eqnarray}
   \hat{C} &\equiv& \frac{C}{4 \pi G r_e^2 \rho_{\max}} \ .
\end{eqnarray}
Here $H_{suffix}$ is the component of the magnetic field 
where $suffix$ may be $c$ (center), $sur$ (surface), 
$p$ ({\it poloidal}) and $t$ ({\it toroidal}). Similarly 
we define normalized global quantities as follows:
\begin{eqnarray}
   \hat{M} = \frac{M}{r_e^3 \rho_{\max}} \ ,
\end{eqnarray}
\begin{eqnarray}
   \hat{W} = \frac{W}{4\pi G r_e^5 \rho_{\max}^2} \ ,
\end{eqnarray}
\begin{eqnarray}
   \hat{T} = \frac{T}{4\pi G r_e^5 \rho_{\max}^2} \ ,
\end{eqnarray}
\begin{eqnarray}
   \hat{\Pi} = \frac{\Pi}{4\pi G r_e^5 \rho_{\max}^2} \ ,
\end{eqnarray}
\begin{eqnarray}
   \hat{U} = \frac{U}{4\pi G r_e^5 \rho_{\max}^2} \ ,
\end{eqnarray}
\begin{eqnarray}
   \hat{{\cal H}} = \frac{{\cal H}}{4\pi G r_e^5 \rho_{\max}^2} \ .
\end{eqnarray}
We also define dimensionless forms of arbitrary functions as 
follows:
\begin{eqnarray}
  \hat{\kappa}(\hat{\Psi}) =
   \left\{
     \begin{array}{lr}
        0 \ , & \mathrm{for} \hspace{10pt} \hat{\Psi} \leq \hat{\Psi}_{\max} \ , \\
        \dfrac{\hat{\kappa}_0}{k+1}(\hat{\Psi} - \hat{\Psi}_{\max})^{k+1} \ , 
         & \mathrm{for} \hspace{10pt} \hat{\Psi} \geq \hat{\Psi}_{\max} \ ,
     \end{array}
   \right.
\end{eqnarray}
\begin{eqnarray}
  \D{\hat{\kappa}(\hat{\Psi})}{\hat{\Psi}} =
   \left\{
     \begin{array}{lr}
        0 \ , & \mathrm{for} \hspace{10pt} \hat{\Psi} \leq \hat{\Psi}_{\max} \ , \\
        \hat{\kappa}_0(\hat{\Psi} - \hat{\Psi}_{\max})^{k} \ , & \mathrm{for} \hspace{10pt} \hat{\Psi} \geq \hat{\Psi}_{\max} \ .
     \end{array}
   \right.
\end{eqnarray}
for $\kappa$ and 
\begin{eqnarray}
   \hat{\mu}(\hat{\Psi})            & = & \hat{\mu}_0 (\hat{\Psi} + \hat{\epsilon} )^m \ , \\
   \int \hat{\mu}(\hat{\Psi}) \, d \hat{\Psi} & = & \frac{\hat{\mu}_0}{m+1}(\hat{\Psi} + \hat{\epsilon} )^{m+1} \ .
\end{eqnarray}
for $\mu$. We choose $k=0.1$, $\hat{\kappa}_0 = 10$ and 
$\hat{\epsilon}= 1.0 \times 10^{-6}$  and keep their values 
fixed during  all calculations in this paper.

\subsection{Computational grids}
\label{App:grid}

We describe the details of our numerical grid points.
In order to resolve the distributions of the source term
of the vector potential equation without loss of accuracy, 
we choose the following non-uniformly distributed grid
points in the actual numerical computations.
In the $\hat{r}$-direction, we divide the whole space 
into two distinct regions: $[0, 1.0]$ (region 1), 
and $[1.0, 2.0]$ (region 2).
In each region, the following mesh points are defined:
\begin{eqnarray}
 \hat{r}_i = w_i^2 \
   \left\{
     \begin{array}{lll}
      \ w_i = (i-1) \Delta w_1 \ ,\
       & \Delta w_1 \equiv \dfrac{\sqrt{1} - \sqrt{0}}{n_1-1} \ , 
       & \mathrm{for} \hspace{10pt} 1 \leq i \leq n_1 \ , \\
      \ w_i = 1.0 + (i - n_1) \Delta w_2 ,\  
      & \Delta w_2 \equiv \dfrac{\sqrt{2} - \sqrt{1}}{n_2-1} ,\
       & \mathrm{for} \hspace{10pt} n_1 \leq i  \ .
     \end{array}
   \right.
\label{Eq:r_mesh}
\end{eqnarray}
where
$n_1$ and $n_2$ are
the mesh numbers defined  as follows:
\begin{eqnarray}
    n_1 & \equiv & \frac{3}{4} (n_r - 1) +1 \ , \\
    n_2 & \equiv & \frac{1}{4} (n_r - 1) +1 \ .
\end{eqnarray}
Here $n_r$ is the total mesh number in the $r$-direction.
In practice, since we use a difference scheme of the 
second-order accuracy for the derivative and Simpson's 
integration formula, we divide each mesh interval defined 
above further into two equal size intervals in the 
$r$ coordinate. We use $n_r = 513$ and thus the actual 
total number of the mesh points is 
$(2 n_r - 1) = 1025$.

\begin{figure*}
\begin{minipage}{150mm}
   \includegraphics[scale=0.6]{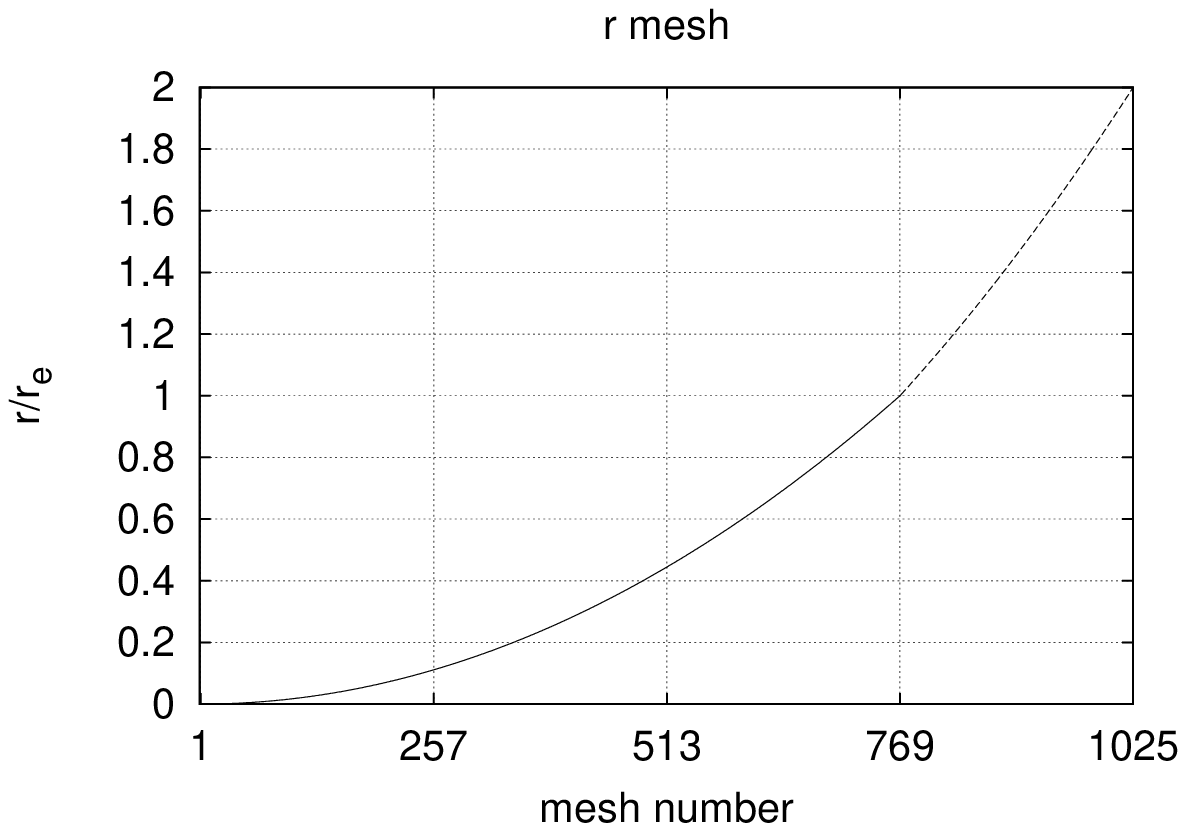}
 \includegraphics[scale=0.6]{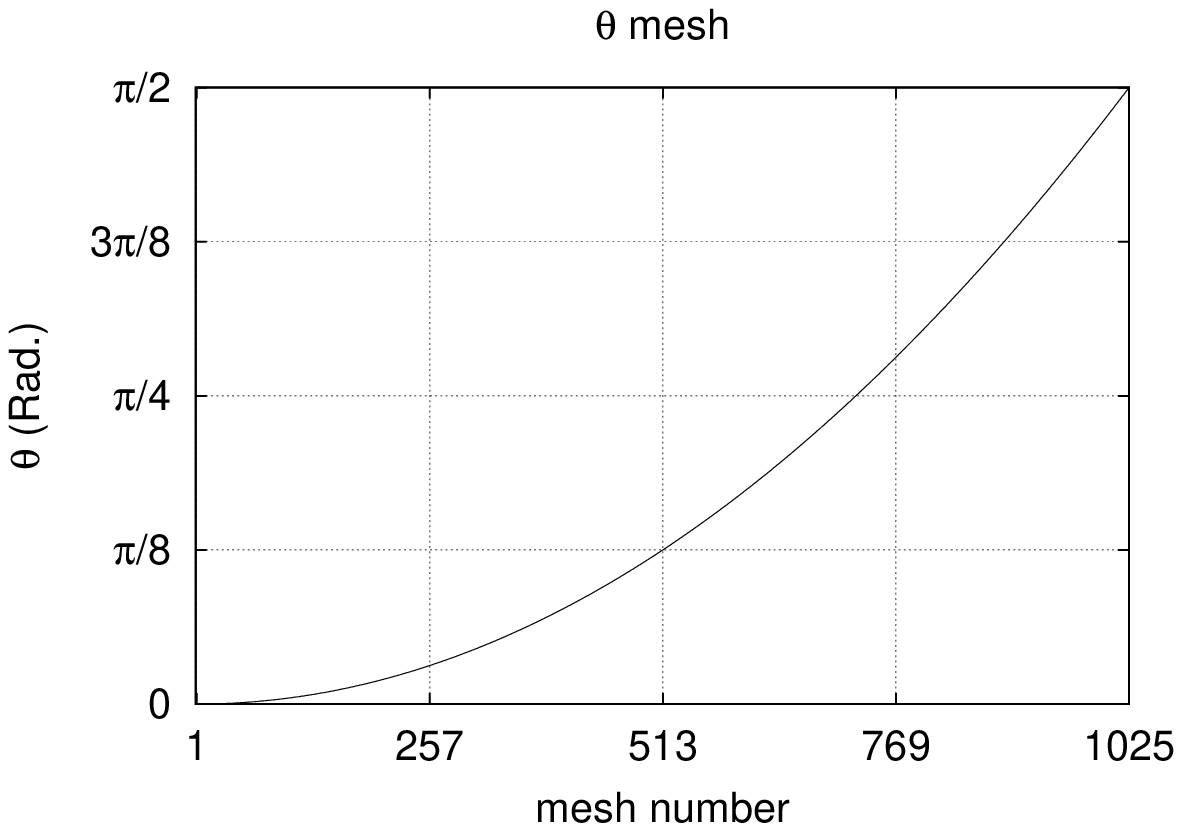}
   \caption{Left: the coordinate $r/r_e$ is plotted as a function
of the grid points. The solid curve shows the region
 1 ($[0,1]$) and the dashed curve shows the region 2 ($[1,2]$).
Right: the same as the left panel except for the $\theta$ coordinate.}
   \label{fig:mesh}
\end{minipage}
\end{figure*}

Concerning the $\theta$-direction, we have to resolve the
region near the axis, because for $m<0$ values the magnetic
fields seem to be highly localized to the axis
region. In order to treat such magnetic fields near the 
axis region, we introduce the following mesh in the 
$\theta$-direction:
\begin{eqnarray}
    \theta_j = \lambda_j^2 \ , \ \lambda_j = (j-1) \ \Delta \lambda \ ,
               \  1 \le j \le n_\theta \ , \  \Delta \lambda = \frac{ \sqrt{\pi /2}}{n_\theta-1} \ ,
\end{eqnarray}
where $n_\theta$ is the total mesh number in the 
$\theta$-direction. We also divide each mesh interval 
defined above further into two equal size intervals. 
Then, we use $n_\theta = 513$ and thus the actual total 
number of the mesh points is $1025$. Fig. \ref{fig:mesh} 
shows the relations between the order of the grid points
and the $r$- or $\theta$-coordinate value.

\end{document}